\documentclass[10pt,a4paper,twocolumn]{article}
\pdfoutput=1
\usepackage[left=10mm,right=10mm,top=20mm,bottom=20mm]{geometry}
\usepackage[utf8]{inputenc}
\usepackage[english]{babel}
\usepackage[T1]{fontenc}
\usepackage{amsmath}
\usepackage{amsfonts}
\usepackage{amssymb}
\usepackage{graphicx}
\usepackage{moreverb,url}
\usepackage{fancyhdr}
\usepackage{lineno}
\usepackage{booktabs}
\usepackage{multirow}
\usepackage{authblk}

\usepackage{natbib}
\usepackage{csquotes}

\usepackage{listings}
\usepackage[colorlinks,bookmarksopen,bookmarksnumbered,citecolor=red,urlcolor=red]{hyperref}
\usepackage[format=plain,indention=5mm, justification=justified, font={small,singlespacing},	labelfont={bf}]{caption}

\author{Marc René Schädler}
\affil{Medizinische Physik and Cluster of Excellence Hearing4all, Universität Oldenburg, Germany\\marc.rene.schaedler@uni-oldenburg.de}





\newcommand{\keywords}[1]{\begin{center}Keywords: #1\end{center}}

\begin{document}
	\rhead{Marc René Schädler}
	\lhead{}
	\date{\today}
  
	\title{Thoughts on the potential to compensate a hearing loss in noise}
	\twocolumn[
	\begin{@twocolumnfalse}
		\maketitle
		\begin{abstract}
        \noindent The effect of hearing impairment on speech perception was described by Plomp (1978) as a sum of a loss of class A, due to signal attenuation, and a loss of class D, due to signal distortion.
        While a loss of class A can be compensated by linear amplification, a loss of class D, which severely limits the benefit of hearing aids in noisy listening conditions, cannot.
        Not few users of hearing aids keep complaining about the limited benefit of their devices in noisy environments.
        Recently, in an approach to model human speech recognition by means of a re-purposed automatic speech recognition system, the loss of class D was explained by introducing a level uncertainty which reduces the individual accuracy of spectro-temporal signal levels.
        Based on this finding, an implementation of a patented dynamic range manipulation scheme (PLATT) is proposed, which aims to mitigate the effect of increased level uncertainty on speech recognition in noise by expanding spectral modulation patterns in the range of 2 to 4\,ERB.
        An objective evaluation of the benefit in speech recognition thresholds in noise using an ASR-based speech recognition model suggests that more than half of the class D loss due to an increased level uncertainty might be compensable.
		\end{abstract}
	\end{@twocolumnfalse}
	]

\section*{Introduction}
\label{sec:intoduction}
To this day, hearing aids without directional amplification or directional noise suppression provide their users only with limited benefit in noisy listening conditions.
The limited benefit of hearing aids in noisy listening conditions is long known and was extensively described and put into context by \cite{plomp1978}.
There, the effect of impaired hearing on speech recognition performance was described as a sum of two fundamentally different classes of hearing loss: class A, which accounts for an attenuation of the signal, and class D, which accounts for a distortion of the signal.
While, the class A loss is defined such that it can be fully compensated by a suitable linear amplification of the signal, the class D loss was assumed to be level independent, that is, it cannot be compensated by linear amplification.

The class A loss, according to \cite{plomp1978} usually makes up the largest part of the total increase in speech recognition thresholds (SRTs) in quiet (A+D), which is why hearing aids provide the largest benefits in quiet environments.
In noisy environments with sufficiently high levels, the contribution of the class A loss diminishes, and the contribution of the class D loss dominates the total loss.
\cite{plomp1978} estimated that, on average, the class A loss accounts for approximately two-thirds and the class D loss for approximately one-third of the total loss, where huge individual variability was expected.
A main drawback of this model is that it does not provide a specific hint on if and how the class D loss can be compensated.

Recently, \cite{kollmeier2016} proposed a modification to the feature extraction stage of the simulation framework for auditory discrimination experiments \citep[FADE,][]{schaedler2016a} to implement a class A loss as an absolute hearing threshold and a class D closs as a level uncertainty.
The simulation approach with FADE employs a re-purposed automatic speech recognition system (ASR) to predict the outcome of a speech recognition test, such as the matrix sentence test \citep{kollmeier2015}.
The FADE approach was already successfully used to predict the outcomes of several speech in noise recognition experiments \citep{schaedler2015,schaedler2016b} as well as the outcomes of basic psycho-acoustic experiments \citep{schaedler2016a} for listeners with normal hearing.
\cite{kollmeier2016} proposed to remove the information that is not available to an individual listener with impaired hearing in the feature extraction stage of the ASR system used in the FADE modeling approach.

\begin{figure}
	\centerline{\includegraphics[width=0.5\textwidth]{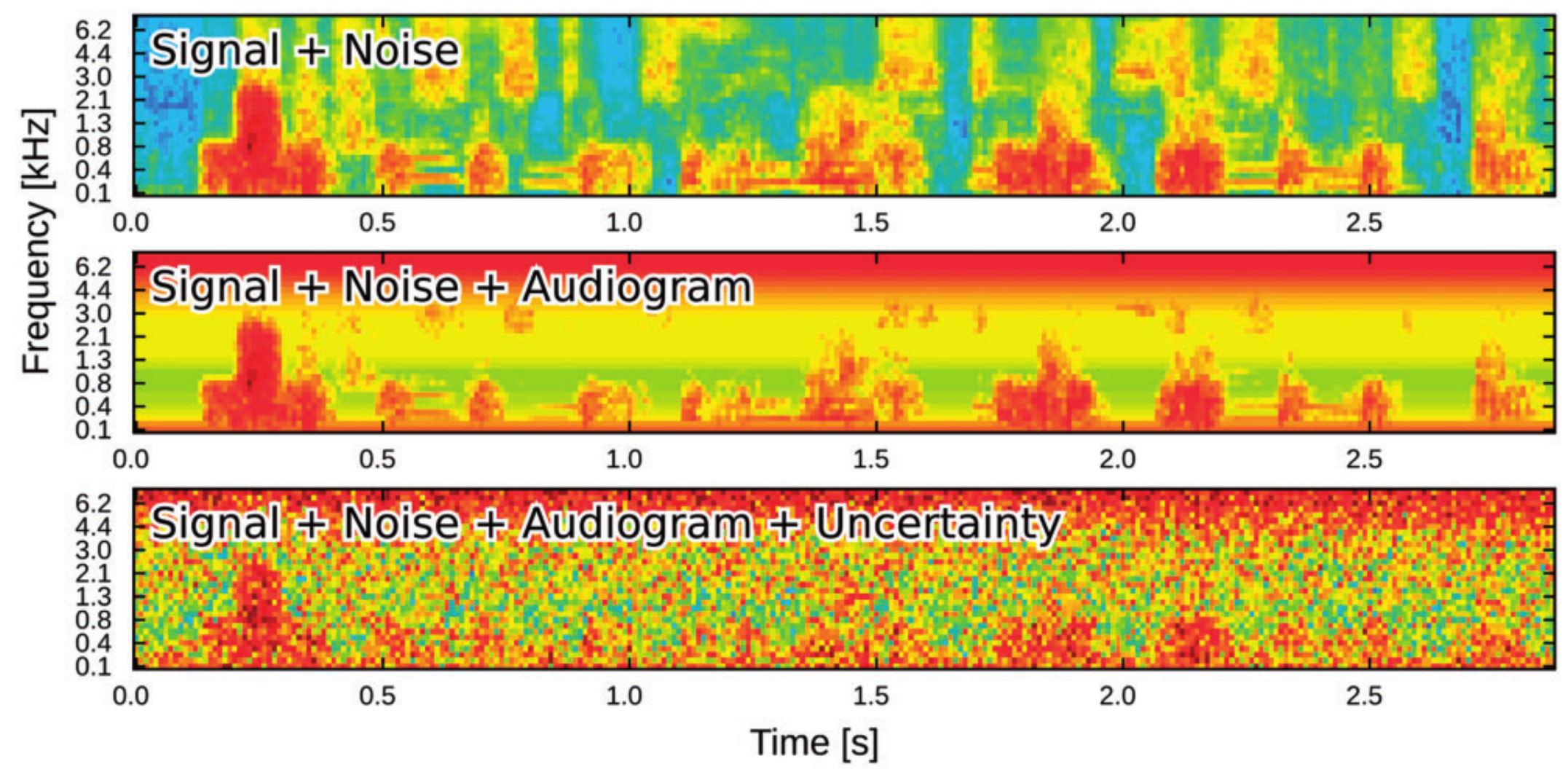}} 
	\caption{Figure reproduced from \cite{kollmeier2015}.
		Internal spectro-temporal signal representation (log Mel-spectrogram) like it is used in the FADE modeling approach of a speech in noise mixture (upper panel) and examples of manipulations to it that were introduced to induce a class A hearing loss (center panel) and a class D hearing loss (lower panel).}
	\label{fig:1}
\end{figure}

To induce a class A loss in the model, variations in the internal spectro-temporal signal levels below the individual hearing threshold, determined by the individual audiogram, were removed.
This manipulation is illustrated in the center panel of Figure~\ref{fig:1}, where the low-energy portions (blue/green) were replaced by constant values which are equal to the individual absolute hearing threshold, while the high-energy portions above the individual absolute hearing threshold are unchanged compared to unmodified representation in the upper panel.
It seems plausible that, if all relevant signal portions are above the hearing threshold, this manipulation should have no effect on the predicted speech recognition performance.

To induce a class D loss in the model, random values were drawn from a normal distribution and added to the internal spectro-temportal signal levels, where the standard deviation of the normal distribution was a variable called \emph{level uncertainty}.
This manipulation is illustrated in the lower panel in Figure~\ref{fig:1}, where all signal portions, including those above the hearing threshold, are affected.
Because the signal energy in that representation (which is a logarithmically scaled Mel-spectrogram) is represented in a logarithmic domain, linear amplification cannot be expected to change its effect on the predicted speech recognition performance. 

\begin{figure*}
	\centerline{\includegraphics[width=0.4\textwidth]{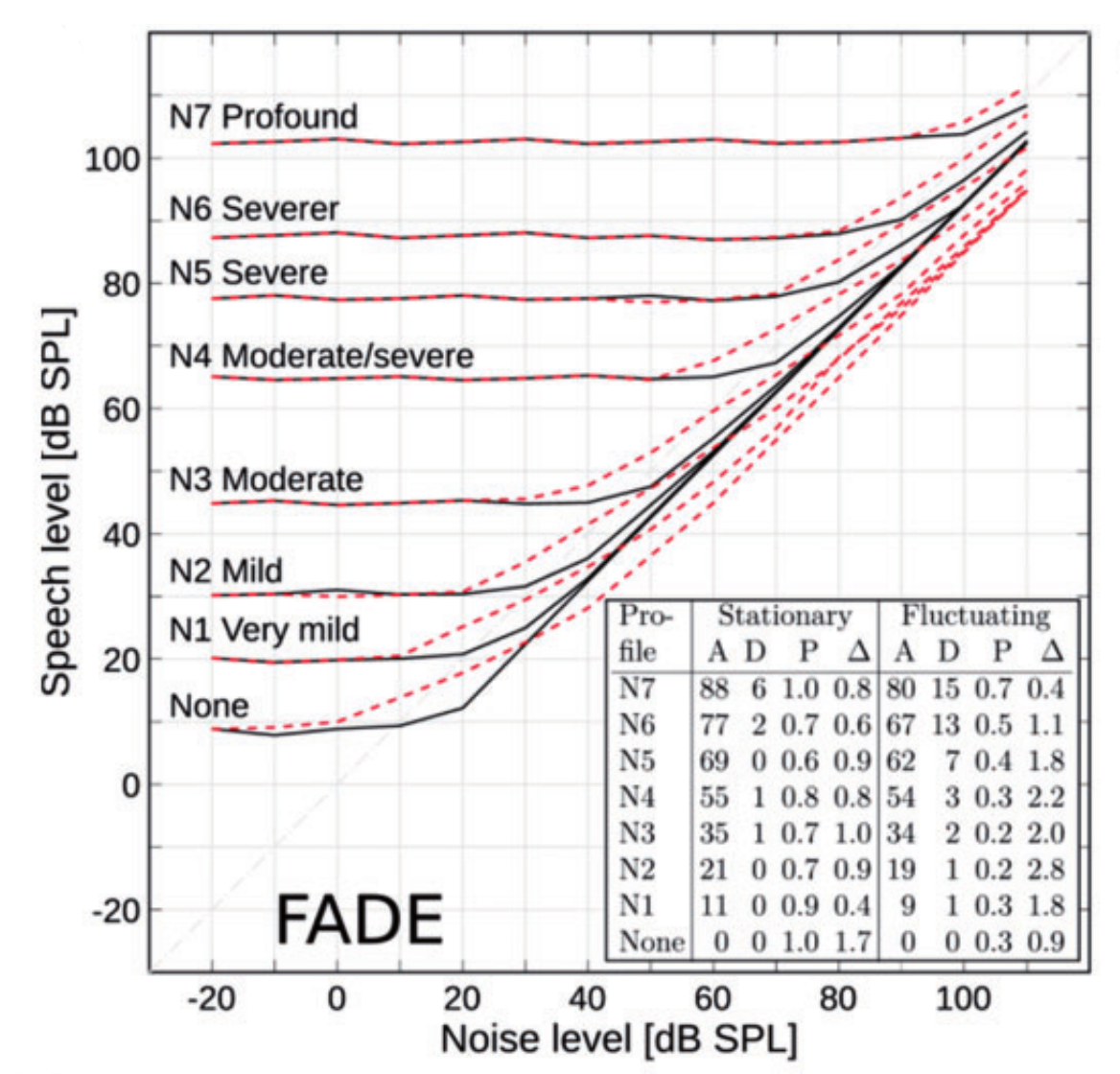}\includegraphics[width=0.4\textwidth]{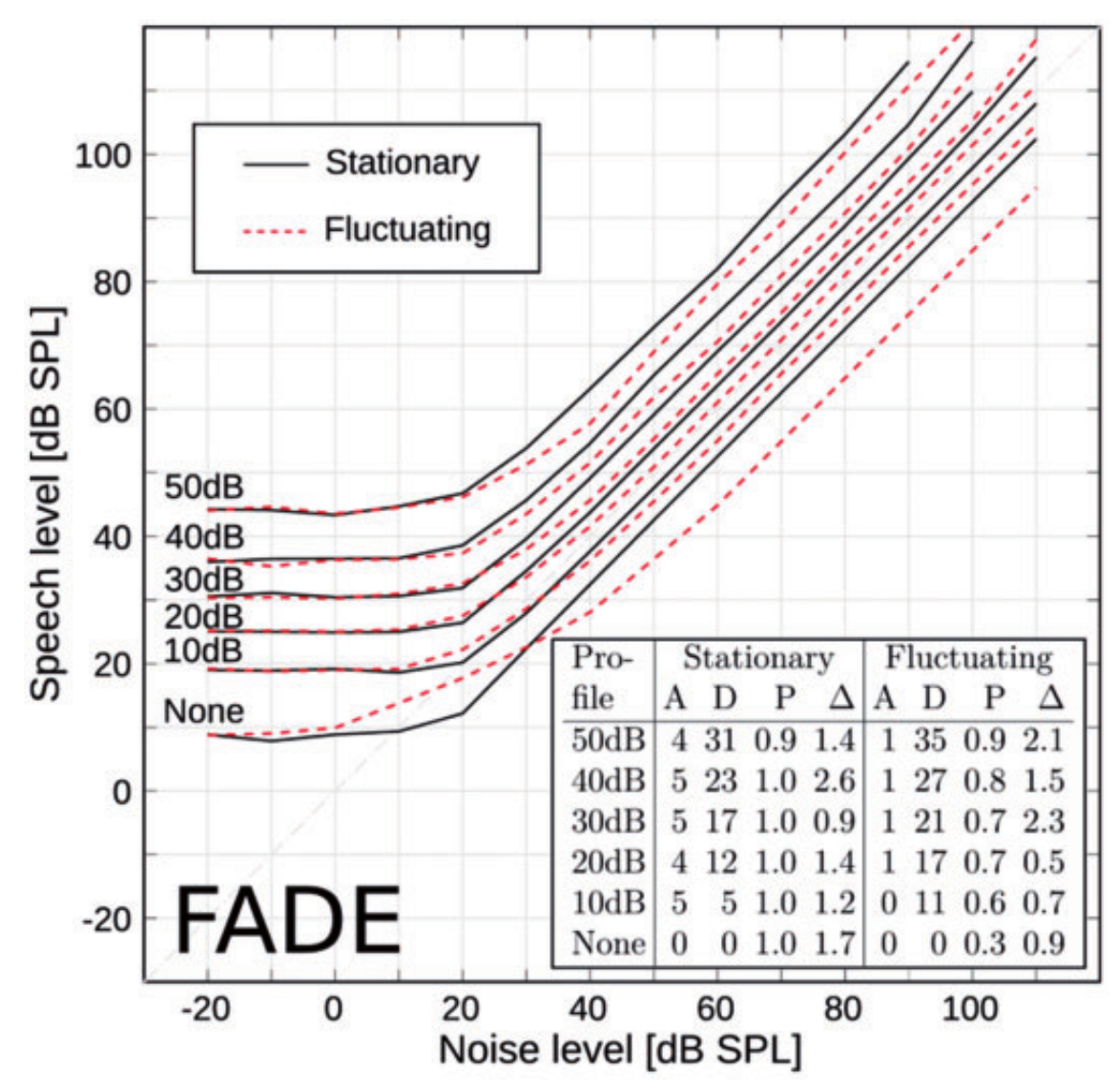}} 
	\caption{Figures reproduced from \cite{kollmeier2015}.
		Simulated speech recognition thresholds with FADE for a stationary and a fluctuating noise condition at different noise levels.
		The left panel shows simulations with different absolute hearing threshold, which induce different class A losses.
		The right panel shows simulations with different values for the level uncertainty, which induce different class D losses.
		The embedded tables show the contributions of A and D when the descriptive model of \cite{plomp1978} was fitted to the data.
		For further details please refer to the original publication.}
	\label{fig:2}
\end{figure*}

\cite{kollmeier2016} evaluated the effect of these manipulations on the predicted outcomes of the German matrix sentence test in a stationary and a fluctuating noise condition for different noise levels, and fitted the A/D-class description proposed by \cite{plomp1978} to the data.
The results, reproduced in Figure~\ref{fig:2}, clearly show that the two manipulations largely achieved the intended effects, that is, inducing a class A and a class D hearing loss.
The left panel shows FADE simulations with different standard audiograms from \cite{bisgaard2010}, which converge for high noise levels, that is, the manipulations \emph{can} be compensated by amplification as one would expect from a class A loss.
The right panel shows FADE simulations with increasing values for the level uncertainty, which do not converge for high noise levels, that is, the manipulations \emph{cannot} be compensated by amplification as one would expect from a class D loss.
An important observation of \cite{kollmeier2015} was, that their empirical data set, which included matrix sentence test results in noise of almost 200 ears, could not be satisfactorily predicted with the class A loss alone, indicating that an implementation of a mechanism that induces a class D loss is needed to explain the speech recognition performance of individual listeners.

\cite{schaedler2020a} extended that approach by inferring the individual frequency-dependent level uncertainty from tone in noise detection thresholds, and achieved unprecedented accuracy in the prediction of benefits in SRTs due to different traditional hearing loss compensation schemes in noise (and in quiet).
The central assumption there is that the same mechanism that affects individual speech in noise perception also affects individual tone in noise perception.
Other mechanisms that reduce the information encoded in the internal signal representation, like a reduced spectral resolution, might also be considered in this modeling approach.
However, \cite{huelsmeier2020} found that, with FADE, a reduced spectral resolution had only little effect on the simulated SRTs compared to an increased level uncertainty.
This indicates, that between a reduced spectral resolution and the level uncertainty, the latter is considered to be the suitable mechanism to implement a class D loss in FADE.

If the mechanism that removes the information in the auditory system of listeners with impaired hearing works similar to the level uncertainty, than the level uncertainty can be seen as the functional counter-part to a compensation strategy for a class D hearing loss.
This is exactly what we will assume in the remainder of this contribution to what we consider theoretical audiology.
In the context of FADE, this will allow to generate testable hypothesis on the achievable benefit in speech recognition performance due to a possible compensation strategy for a class D loss.
The aim of this contribution is to:
\begin{itemize}
	\item[A)] Present an approach which is able to partially compensate a class D loss as implemented with the level uncertainty in FADE, and
	\item[B)] objectively evaluate this approach and come up with testable quantitative hypothesis on the benefit in noisy listening conditions.
\end{itemize}

For an effective mitigation of the effect of the level uncertainty on speech recognition performance, three main problems need to be addressed:
\begin{itemize}
  \item[1)] Which portions, or more precisely, patterns of the speech signal carry the most relevant information and \emph{need} to be protected?
  \item[2)] Which signal patterns \emph{can} be protected, given the strict constrains on the available future temporal context of the signal (approximately 1\,ms) in hearing aid applications?
  \item[3)] How can such a protection be achieved?
\end{itemize}

Continuing to follow an (assumed) analogy of basic principles in human and machine speech recognition, literature on robust automatic speech recognition provides hints on which signal patterns are relevant for good (automatic) speech recognition performance in noise.
Let us assume that a log Mel-spectrogram, such as it used for the calculation of the widely used Mel-frequency cepstral coefficient (MFCC) features, is representative for the information that is available to an ASR system.
The spectral resolution of such a log Mel-spectrogram, of which an Example is depicted in the upper panel of Figure~\ref{fig:1}, is about 1 equivalent rectangular bandwidth (ERB), the temporal resolution is 10\,ms.
The relevant speech information is encoded in the represented spectro-temporal dynamic, that is, the differences of spectro-temporal signal levels over time, called temporal modulations, and over frequency, called spectral modulations.
It is remarkable that ASR systems traditionally don't even use the whole information in the log Mel-spectrogram, but work with a reduced spectral resolution compared to the spectral resolution of the human auditory system.
For example, \emph{the} standard features used for ASR, MFCCs \citep{etsi2007}, specifically encode spectral modulation frequencies from $0$ to about $\frac{6}{23}$ cycles per ERB; spectral modulation frequencies above $\frac{1}{4}$ cycles per ERB empirically don't contribute to automatic speech recognition performance.
In line with this finding, the robust Gabor filter bank (GBFB) features use only slightly more than half of the available spectral resolution of the log Mel-spectrograms \cite[c.f.][]{schaedler2012}.
By omitting the corresponding parts of the feature vector, \cite{schaedler2012} assessed the relative importance of different spectro-temporal modulation frequencies in a robust ASR task.
There, it was observed that the highest spectral modulation frequencies, around $\frac{1}{4}$ cycles per band (the bandwidth of a Mel band there is approximately 1 ERB) seem to important for the considered speech in noise recognition task.
Before going into more detail on the importance of spectro-temporal modulations for noisy speech recognition,
the limited possibilities to manipulate these in the context of hearing aids have to be considered.

The restriction to the availability of approximately 1\,ms future temporal context in hearing aids does not allow for the reliable  manipulation of temporal modulations below approximately 500\,Hz.
This limit is still way above the temporal modulation frequencies that are represented in the features of ASR systems.
With the common analysis window length and shift of 25 and 10\,ms, respectively, as used in Figure~\ref{fig:1}, the upper limit for represented temporal modulation frequencies is around 50\,Hz.
Hence, the only modulations that can be reliably manipulated are spectral modulations, which requires a temporal signal context in the order of 1\,ms.
Fortunately, spectral modulations seem to play an important role in (automatic) speech recognition.
With the already mentioned approach to model human speech recognition performance by means of a re-purposed ASR system, \cite{schaedler2016a} found that explicitly encoding spectral modulations, that is, across-frequency interactions, in the feature vector was required to explain the empirically found benefit when listening in a fluctuation masker.
In other words, spectral modulations seem to be especially relevant in fluctuating noise maskers.

In combination of the answers to problems 1) and 2): Spectral modulations just below 0.25 cycles per ERB, let's say spectral patterns between 2 and 4 ERB, seem to be a good first candidate for protecting them against the level uncertainty.
How can these signal patterns be protected or at least hardened against the effect of the internal noise due to the level uncertainty?
In the logarithmic level domain (in the log Mel-spectrogram) the noise of the level uncertainty is additive, and one effective way of mitigating the effect of an additive noise is amplification.
This means, that the desired spectral modulation patterns should be amplified, that is, expanded before the noise of the level uncertainty can remove that information.

At a first glance, an expansion of signal dynamic in the context of the reduced residual dynamic range of listeners with impaired hearing might seem completely undesirable:
The expansion of spectral modulations can result in uncomfortably high and/or inaudibly soft signal levels; which can occur at the same time in different frequency ranges.
However, considering the scale (2 to 4 ERB) it is clear, that this is a region which is not modified by common approaches to multi-band dynamic compression, where the signal is usually independently compressed in approximately six bands.
The traditional approaches to multi-band dynamic compression with less than 9 bands leave spectral patterns with a width of 2 to 4 ERB virtually uncompressed, which might also be interpreted as an indication towards the importance of these patterns. 
If now the expansion of one (small, as we will later see) part of the spectral dynamic increases the total spectral dynamic range of the signal, it might be possible to counter it with a stronger compression of other (less relevant) parts of the spectral dynamic.
However, compression should only be applied if it is required, that is, if the input signal dynamic range does not fit into the available output dynamic range.
The dynamic range of speech in noise is less than the dynamic range of a clean speech signal.
Especially when considering speech in only slightly fluctuating noises at signal-to-noise ratios (SNRs) around 0\,dB, the dynamic range of the mixture can be close to the minimum which is required to discriminate words.
In this condition, no improvement can be expected from compressing the signal dynamic.
Hence, a micro-management of the limited residual dynamic range to optimize its use for speech-recognition-relevant information is a desirable goal.
PLATT is an approach which gives preference to spectral modulation patterns that are supposedly relevant for speech recognition at the expense of spectral modulation patterns which are supposedly less relevant when mapping an input signal dynamic to a given output signal dynamic.
The here outlined ideas are part of a German patent (DE 10 2017 216 972); a patent application in Europe is pending.

The implementation details of PLATT, which are presented in the Section~\nameref{sec:methods}, were already engineered towards low-delay real-time processing.
In a simpler implementation, the representation of the log Mel-spectrogram could be manipulated in the MFCC domain where the spectral modulations are suitably encoded for this operation, and which in MFCC-terminology would be called \emph{liftering}.
The inverse transform of the modified MFCCs, with amplified spectral modulation patterns between 0.25 and 0.125 cycles per band, could be compared to the unmodified log Mel-spectrogram, and a suitable filter response in the time-domain could be designed for each signal frame.
While such an approach would have the advantage of simplicity (even simpler than the related implementation of an intelligibility-improving signal processing approach (IISPA) \cite{schaedler2020b}), it would not be applicable without further modifications in a hearing device: The main problem would be the infeasible window length of 25\,ms.
Because of the highly non-linear nature of the interactions between the factors influencing speech recognition performance (speech material, masker type and level, reverberation, non-linear signal processing, hearing impairment), an implementation that already fulfills the most basic requirements of a hearing aid algorithm and can run on a hearing aid prototype\footnote{\url{https://github.com/m-r-s/hearingaid-prototype}} was preferred over a simple proof-of-concept implementation.
This additional effort makes a seamless translation to an application in a hearing device more likely and increases the meaningfulness of the presented results for a possibly realizable hearing aid solution.

For an evaluation of the implementation with respect to a possible compensation of a class D loss, the following points need to be considered:
\begin{itemize}
  \item[1)] Which listening conditions, that is, which speech test and maskers, are suitable to evaluate the PLATT implementation objectively with FADE \emph{and} (also later) empirically.
  \item[2)] Which listener profiles are suited to clearly demonstrate a (partial) compensation of a class D loss like it is implemented in FADE.
\end{itemize}

The first point is important to enable the verification with empirical data of any hypotheses that are based on the model predictions.
\cite{schaedler2020a} discussed this point and proposed to use the SRT-50 measured with the matrix sentence test in quiet, in a stationary, and in a fluctuating noise condition, to cover the very different masker properties in typical listening conditions: Quiet, low-dynamic maskers, high-dynamic masker.
The main reason for using these \enquote{laboratory} signals and the SRT-50 instead of real noise recording and, e.g., the SRT-80, was the known high test-retest reliability of these tests.
For individual predictions, the test-retest reliability of the employed speech recognition test sets the lower limit for the achievable prediction error in an evaluation with the data.
That means, low measurement errors in the empirical data may facilitate or even enable falsification of the model predictions.
An SRT of 0\,dB at high noise levels in the test-specific noise condition, this is, with a stationary noise of identical long term spectrum than the speech signal, can be considered very problematic when normal-hearing listeners can achieve about -8\,dB.
If only half of the hearing loss in that condition in noise could be compensated (which would be a huge achievement), the measurable benefit would be only 4\,dB.
Considering that the benefit in SRT is calculated as the difference between two measurements, the targeted error of a single measurement should be less than $\frac{1}{\sqrt{2}} \cdot \frac{1}{2} \cdot 4\text{dB} \approx 1.41\,\text{dB}$.
Such low measurement errors, which can be achieved in SRT measurements with the matrix sentence test, would later enable to show individual benefits without averaging over groups of listeners, given the benefit was 4\,dB.
When adding to this consideration the need for a level-dependent evaluation, which is required to identify the class D loss according to \cite{plomp1978}, one arrives at the test conditions which were already studied in \cite{kollmeier2015} and that are depicted in Figure~\ref{fig:2}.

The second point, the selection of suitable listener profiles, is a bit more complex than it might initially appear.
A sensible approach would be to take the individual profiles inferred from the psychoacoustic measurements by \cite{schaedler2020a} which are available online\footnote{\url{https://doi.org/10.5281/zenodo.4394186}}.
The main problem with this approach is, that even with a small pure class A loss, the SRTs are generally not level-independent at high levels.
This can already be observed for the fluctuating noise condition in the left panel of Figure~\ref{fig:2}.
There, the simulations with a pure class A loss with hearing thresholds according to the standard profile \emph{N1} (corresponding to a very mild hearing loss) do not converge with the data of the normal-hearing profile \emph{None} up to noise levels of 90\,dB SPL.
Hence, even for very small increases in hearing threshold, amplification improves the SRT in the fluctuating noise condition up to very high presentation levels.
With the aim of clearly attributing compensation strategies to compensate a class A or a class D loss, this is highly undesirable.
To clearly identify the compensation of a class D loss, simple linear amplification alone must not improve the SRT.
The reason for the observed model behavior is that, for the Bisgaard profiles, the frequency range above the hearing threshold increases with the presentation level.
While the limited frequency range can safely be assumed a factor contributing to a class D loss and has to be considered in a suitable listener profile, it must be avoided that the effectively used frequency range changes at high presentation levels.
One option would be to low-pass filter the speech material.
Another option is to define profiles with very steep sloping hearing loss functions.
The former option would be very suitable to measure empirical data.
The latter option is regarded cleaner from a modeling perspective, because it reduces the number of parameters that influence the SRT and results in a simpler and possibly better traceable model.
Hence, listener profiles with normal hearing thresholds below, and infinite hearing loss above a given limit frequency are suitable for the considerations in this contribution.
While the measurements from \cite{schaedler2020a} indicate that the level uncertainty might be frequency-dependent, profiles with frequency-dependent level uncertainty would add an additional dimension to a already complex matter.
Also, the data in that study was from a non-representative group of 18 listener of which it is unknown if extreme cases of class D loss were present.
Hence, for the purpose of demonstrating the effectiveness of the compensation approach with PLATT and to discover its interactions and limits, idealized profiles that can be described by only two parameters, the frequency limit and (frequency-independent) level uncertainty, are preferable.
In later measurements with human listeners, the limited frequency range can be achieved by low-pass filtering the signals to match the considered conditions.

With these considerations the following steps are enabled:
\begin{itemize}
	\item The applicable functional description of a process that is assumed to cause a class D hearing loss, the level uncertainty, can be used to design a matched functional compensation strategy.
	\item Simulation experiments with FADE can be conducted which quantify the compensation of a class D loss as implemented with the level uncertainty as a function of frequency range and level uncertainty.
	\item These experiments can later also be conducted with human listeners to test the hypothesis.
\end{itemize}

\section*{Methods}
\label{sec:methods}
The methods described in the following were used to simulate speech recognition experiments in stationary and fluctuating noise at different presentation levels for 16 listening profiles with class D hearing losses without and with the later proposed dynamic range expansion by PLATT including different degrees of expansion.

\subsection*{Speech recognition tests}
\label{sec:matrixtests}
\begin{figure*}
	\centerline{\includegraphics[width=1.0\textwidth]{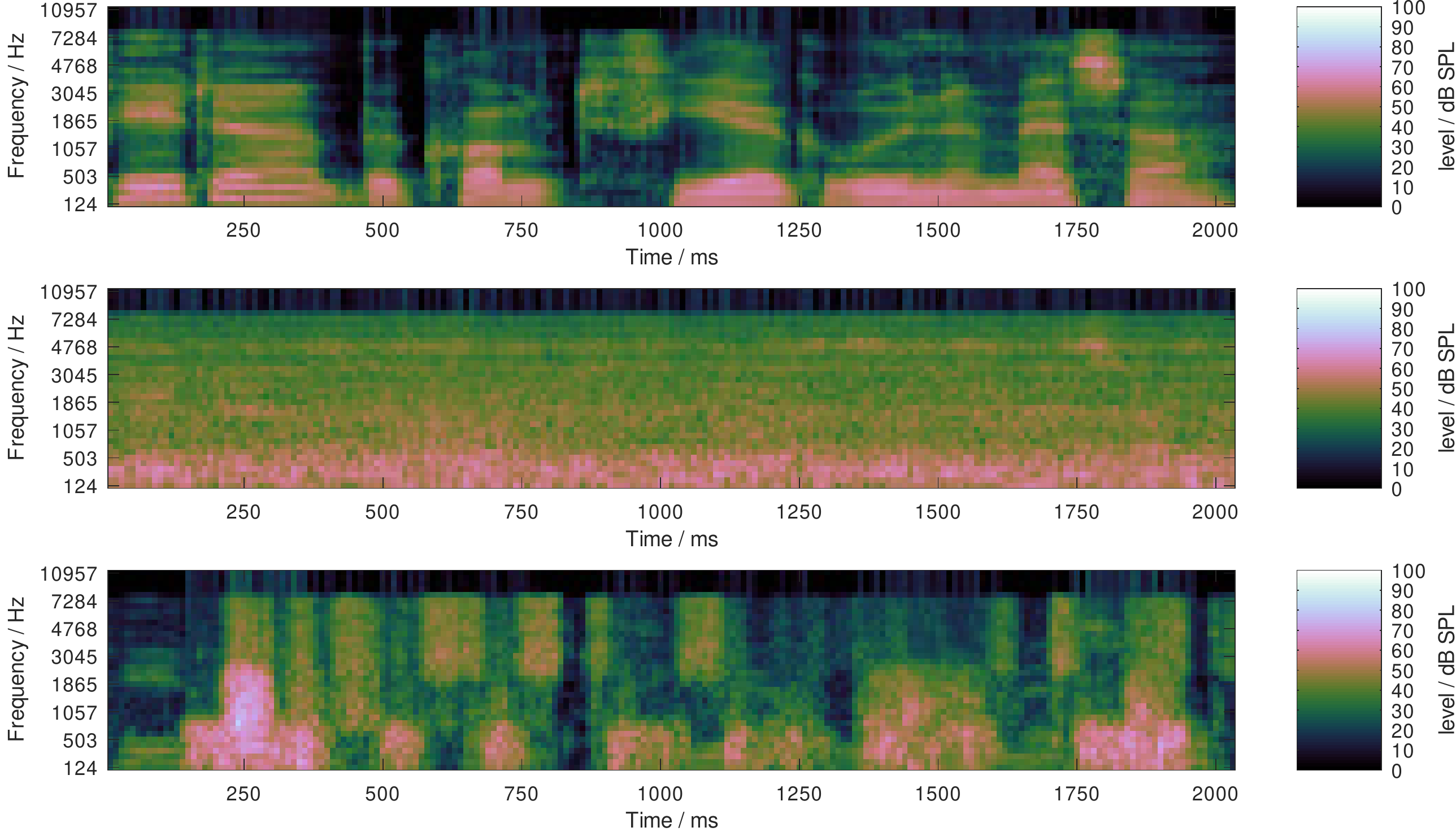}}
	\caption{Log Mel-spectrograms of a clean German matrix sentence at 65\,dB SPL (upper panel), of the same sentence in the stationary noise (center panel) and fluctuating noise (lower panel) conditions at the SNRs which correspond to the SRT listeners with normal hearing of speech, -7 and -19\,dB, respectively.}
	\label{fig:3}
\end{figure*}

The speech material of the (male) German matrix sentence test \citep{wagener1999,kollmeier2015} was used with two masker signals: The test-specific noise (called OLNOISE) and the fluctuating ICRA5-250 noise signal \citep{dreschler2001,wagener2006}.
The matrix test, which exists in more than 20 languages, comprises 50 phonetically balanced common words of which sentences with a fixed syntax, such as \enquote{Peter got four large rings} or \enquote{Nina wants seven heavy tables}, are built.
Typically, the SRT-50, this is the speech level that is required to correctly recognize 50\% of the words, is measured with an adaptive procedure in experiments with human listeners.
In this contribution, the speech and masker material was used to predicted the SRT-50 with FADE.
The test-specific noise (OLNOISE) has the same long-term spectrum as the speech material.
It can be assumed to mask the speech signal similarly well across all frequencies.
At the SRT for normal hearing listeners, -7\,dB \citep{hochmuth2015}, this results in a noisy speech signal with a low dynamic range, where spectro-temporal maxima of the mixtures are dominated by the speech signal.
The effect can be observed in Figure~\ref{fig:3}, where the log Mel-spectrogram of clean speech signal (upper panel) and the same speech signal with the OLNOISE masker at -7\,dB SNR (center panel) are depicted.
The ICRA5-250 noise is a speech-shaped noise which is co-modulated with speech-like temporal patterns in three independent frequency bands, where the pause duration was limited to 250\,ms \citep{wagener2006}.
The empirical SRTs with this masker signal are usually more than 10\,dB lower than in the corresponding test-specific noise condition \citep{hochmuth2015}.
At the SRT for listeners with normal hearing, -19\,dB \citep{hochmuth2015}, this results in a noisy speech signal with a high dynamic range, where the spectro-temporal maxima of the mixtures are dominated by the masker signal.
This can be observed in the lower panel of Figure~\ref{fig:3}, where the log Mel-spectrogram of a speech signal with the ICRA5-250 masker at -19\,dB SNR is depicted in the lower panel.
To assess the presentation level dependency, both maskers are considered at presentation levels from 0 to 100\,dB SPL in 10-dB steps.
At 0\,dB SPL presentation level, this corresponds to listening in quiet.
The selected listening conditions reflect important dimensions of speech perception: Listening in quiet, at low levels, and at high levels, as well as listening in stationary and fluctuating noise.
All considered speech tests can also be performed human listeners.

\subsection*{Simulations of matrix tests with FADE}
\label{sec:fade}
The speech tests considered in Section~\nameref{sec:matrixtests} were simulated with an ASR-based approach and their outcome, the SRT-50, was predicted based on the simulation results.
The simulations were performed with the latest standard version of FADE\footnote{\url{https://doi.org/10.5281/zenodo.4003779}}, as described by \cite{schaedler2016a}.
In this contribution, FADE is used as a tool to predict the outcome of speech recognition tests, where the only change to the standard setup was the manipulation of the feature extraction stage which is explained in Section \nameref{sec:listenerprofiles} and the processing of the noisy speech signals with PLATT as explained in Section \nameref{sec:platt}.
Hence, the FADE simulation method is only outlined here, and we refer the interested reader to the original description from \cite{schaedler2016a}.

Predictions with FADE are performed completely independently for each listening condition (masker, maskerlevel, hearing loss compensation, and hearing profile).
There is no dependency on any empirically measured SRT, nor on predictions of the same model in other/reference conditions (and hence no need to define such).
This means, with FADE, a single SRT of a speech recognition test for which no empirical data exists can be predicted.
For the prediction of one outcome the following standard procedure as described by \cite{schaedler2016a} was used.

An ASR system was trained on a broad range of SNRs with noisy speech material form the considered condition, e.g. German matrix sentence test in OLNOISE for listener profile \enquote{P-8000-1} with no compensation.
For this, a corpus of noisy speech material at different SNRs was generated from the clean matrix sentence test material and the masker signal, by adding randomly chosen masker signal fragments with the speech material.
The noisy signals were processed with PLATT when an aided listening condition was considered.
From the noisy (and optionally processed) speech signals, features were extracted, where this step included the implementation of the class D hearing loss, as described in Section~\nameref{sec:listenerprofiles}.
Subsequently, an ASR system using whole-word models implemented with Gaussian Mixture Models and Hidden Markov Models, was trained on the features.
This resulted in 50 whole-word models for each training SNR.
These models were then used with a language model that considers only valid matrix sentences (of which $10^5$ exist) to recognize test sentences on a broad range of SNRs with noisy speech material form the same considered condition.
For each combination of a training SNR and a test SNR, the transcriptions of the test sentences were evaluated in terms of the percentage of correctly recognized words.
The resulting recognition result map (cf. left panel in Figure~7 in \cite{schaedler2016a} for an example), which contained the speech recognition performance of the ASR system depending on the training and testing SNRs in 3\,dB steps, was queried for the SRT.
For a given target recognition range, e.g. 50\%, the lowest SNR at which this performance was achieved was interpolated from the data in the recognition result map and reported as the predicted SRT for the considered condition.
The whole simulation process, including the creation of noisy speech material, an optional processing of this noisy speech material with PLATT, the feature extraction (which depends on the listener profile), the training of the ASR system, the recognition of the test sentences, and the evaluation of the recognition result map, was (independently) repeated for each considered condition.

\subsection*{Listener profiles: Class D hearing losses}
\label{sec:listenerprofiles}
\begin{figure*}[h]
	\centerline{\includegraphics[width=1.0\textwidth]{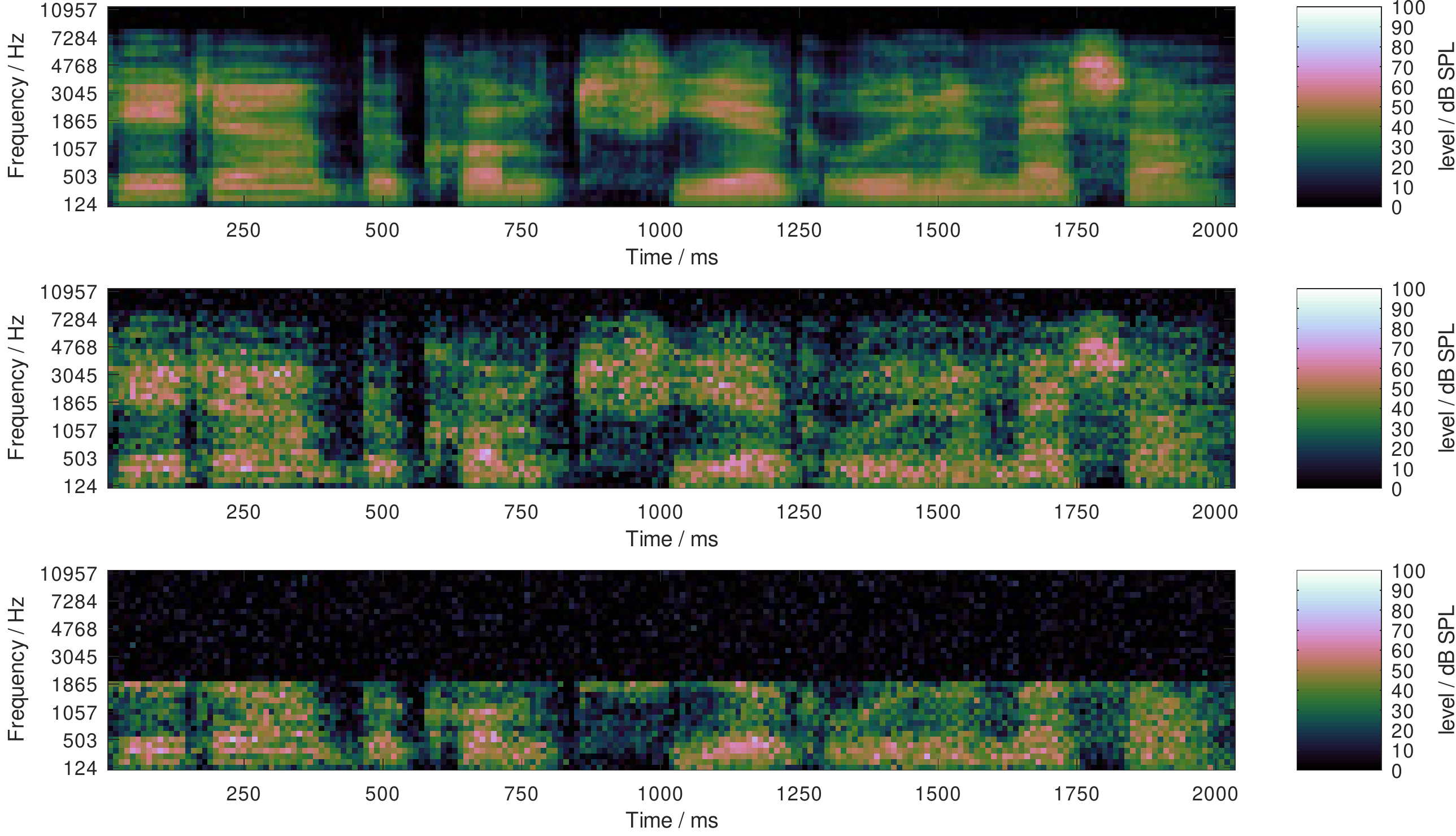}}
	\caption{Illustration of the two considered class-D-loss-inducing log Mel-spectrogram manipulations: Log Mel-spectrograms for listener profiles \enquote{P-8000-1} (upper panel), \enquote{P-8000-7} (center panel), and \enquote{P-2000-7} (lower panel).
	The first number in the profile encodes the upper frequency limit, in this example 8000 and 2000\,Hz, and the second number indicates the level uncertainy, here 1 and 7\,dB.}
	\label{fig:4}
\end{figure*}

Outcome predictions of speech recognition tests as well as basic psychoacoustic tests with FADE were found to be close to the empirical results for listeners with normal hearing \citep{schaedler2016a}.
As proposed by \cite{kollmeier2016} and successfully used by \cite{schaedler2020a}, impaired hearing was implemented in the ASR system by removing the information from the feature vectors that is presumably not available to listeners with impaired hearing.
As discussed in the \nameref{sec:intoduction}, two types of manipulations which induce class D hearing loss were considered: 1) A limitation of the frequency range, 2) An increase of level uncertainty.
The effect of both parameters on the log Mel-spectrogram of a clean speech sample is depicted in Figure~\ref{fig:4}, where in the upper panel, the frequency range was limited to 8000\,Hz and the level uncertainty was 1\,dB.
In the center panel, the level uncertainty was increased to 7\,dB, compared to the upper panel.
In the lower panel, the frequency range was additional limited to 2000\,Hz compared to the center panel.
An amplification of the input signal increases all values in a log Mel-spectrograms by a constant value.
Both manipulations introduce a level-independent loss of information and hence induce a class D loss.
For the evaluation, upper frequency limits of 1000, 2000, 4000, and 8000\,Hz were considered, where the class D loss decreases with high values.
For the level uncertainty, frequency-independent values of 1, 7, 14, and 21\,dB were considered, where the class D loss increases with high values.
All combinations of both parameters result in 16 profiles from \enquote{P-8000-1} to \enquote{P-1000-21}.
The former can be expected to be the best performing one, while the latter can be expected to be the worst performing one.

\subsection*{PLATT dynamic range manipulation}
\label{sec:platt}
In this section, the patented (DE 10 2017 216 972) PLATT dynamic range manipulation as it was conceived for a later implementation in a hearing device is described.
The implementation was optimized to run in real-time on a Raspberry Pi 3 Model B to enable field studies with mobile hearing aid prototype hardware\footnote{For example: \url{https://github.com/m-r-s/hearingaid-prototype}\\ or \url{https://batandcat.com/portable-hearing-laboratory-phl.html}}.
The ability to expand spectral modulation frequencies in the range of $\frac{1}{8}$ to $\frac{1}{4}\frac{\text{cycles}}{\text{ERB}}$ is a feature that integrates naturally with the approach.
Even if not strictly necessary for the goals of this contribution, the method is described here in detail to make statements about its ability to compensate a class D loss in the algorithmic context in which it might be later usable in a hearing device.
To motivate the design decisions behind PLATT, which generally aims to preserve relevant speech modulations when compressing the dynamic range of a signal, this subsection comes with its own introductory part.

\paragraph{Introduction to the PLATT concept}
Conditions in which the available dynamic range for acoustic communication is reduced are rather the norm than the exception.
For example, in a driving car, the lower limit of the available dynamic range is determined by the driving noise.
Or in a library, the upper limit is given by the accepted sound levels in such an environment.
And, importantly, the available dynamic range for communication is limited for listeners with impaired hearing.
For a successful communication, it may be required to adapt a source signal, which may contain speech and non-speech parts, to the available dynamic range on the receiver side by dynamic range compression.
But, in many real-time applications, the available temporal context to perform this operation is very limited.

Multi-band dynamic range compressors used in hearing (aid) research \cite[e.g. ][]{grimm2015} statically map the input dynamic range to a reduced output dynamic range in a number of independent frequency bands.
The compression is often applied with rather short attack time constants, e.g., 20\,ms, with the aim to protect the user from high levels, while the release time constants are usually much longer, e.g. 100\,ms to 1000\,ms, with the aim to limit compression when it not desirable, i.e. during short speech pauses.
However, no distinction is made whether the signal contains speech portions or not.
Approaches which depend on a classification whether or not speech is present in the input signal, are prone to errors if the (speech-)signal-to-noise ratio (SNR) is low, that is, just when the classification result is most important.
Approaches that require more than a few milliseconds of future temporal context cannot be used in applications which require low latency, such as, e.g., hearing devices.
Regarding the speech intelligibility of processed signals, compression in a few wide frequency bands is preferred over compression in many narrow frequency bands, however, the recommended number of channels greatly varies (usually between 1 and 8) \citep{plomp1988,dreschler1992,hohmann1995,yund1995,moore1999,souza2002}.
The fewer channels are used, the better the spectral dynamic, that is, the spectral contrast or \emph{spectral modulation}, is preserved.
Static dynamic range compression only preserves the spectral modulation within each independent frequency band, but not across bands, even if the dynamic range would be available.
Also, fewer frequency channels reduce the need for sharp filters which would require long integration time constants and introduce additional latency.

\cite{bustamante1987} proposed to compress the first two principle components (PC1 and PC2) of the short-term speech spectrum, which were roughly representative of overall level and spectral tilt.
With this approach, the frequency bands were not processed independently anymore, and the finer spectral structure was always preserved.
Their analysis indicated that the highest intelligibility was obtained when audibility was improved and the relative spectral shapes of different speech sounds were preserved \citep{bustamante1987}.
In their concluding section, they recommended to investigate the enhancement of spectral differences while compressing level variations.
\cite{levitt1991} proposed an approach which decomposes and manipulates the short-term spectrum using a set of orthogonal polynomial functions with the aim to preserve important speech cues.
Referring to the study of \cite{bustamante1987}, \cite{levitt1991} wrote: \emph{\enquote{Both studies showed that compression of the lowest order component (factor 1 in the principal-components method and the constant term in the orthogonal polynomial method, respectively) had by far the largest effect, and that compression of higher order components had little effect, if any.}}
The common idea behind these two studies was to linearly map and manipulate the spectral dimension of a suitable spectro-temporal representation with the aim of separating important from less important speech signal dynamic.
However, both studies considered only clean speech signals, and hence did not consider the relevant portions of speech signal for their recognition in noise.

That the signal dynamic can be described as the difference of frequency-dependent short-term effective amplitudes, e.g., across time (temporal dynamic), across frequency (spectral dynamic), or both (spectro-temporal dynamic), raises the question which representation is most suitable to manipulate it.
ASR systems are \emph{the} technical solution to decode speech signals and hence provide a model for speech recognition.
As outlined in the \nameref{sec:intoduction}, the feature extraction stages of ASR systems provide representations of the speech dynamic that are well suited for the robust recognition of speech in noise.
However, the often employed basis for the feature extraction stages, the log Mel-spectrogram, is not suited for low-latency signal processing due to its long integration window.
The relatively long integration window of the log Mel-spectrogram serves two objectives: 1) Obtain a sufficiently high frequency resolution to separate low-frequency signal content into approximately 1\,ERB-wide bands, and 2) Ensure that in voiced speech portions each signal frame contains at least one pulse (that is to remove the temporal fine-structure of the speech signal).
Fortunately, these two aspects (sufficient spectral resolution for low frequencies and limited temporal resolution) are compatible and can be optimized for low-latency processing at the cost of a frequency-dependent group delay.
In the following, the design of PLATT, a fast adaptive dynamic range manipulation scheme that takes the mentioned observations into account, is proposed, where the following three objectives were pursued:
\begin{itemize}
	\item Preservation and enhancement of spectral modulations which are assumed to be relevant for speech recognition
	\item Low-latency and fast reaction time while minimizing audible artifacts
	\item Adaptive limitation of the compression to the necessary minimum
\end{itemize}
PLATT consists of three functional parts:
\begin{itemize}
	\item[1)] An auditory-motivated frequency decomposition and re-synthesis of the audio signal, which allows to manipulate frequency- and time-dependent amplitudes in a perceptually relevant domain and helps to ensure that only limited audible artifacts can be introduced.
	\item[2)] Extraction of a spectro-temporal representation from the frequency-decomposed signal which is similar to those used for robust ASR and hence suitable for the analysis of the relevant spectral modulations.
	\item[3)] Adaptive calculation of frequency- and time-dependent gains from the spectro-temporal representation which uses compression only as required to provide high speech recognition performance when the available dynamic range on the output side is limited.
\end{itemize}

\begin{figure*}
	\centerline{\includegraphics[width=1.0\textwidth]{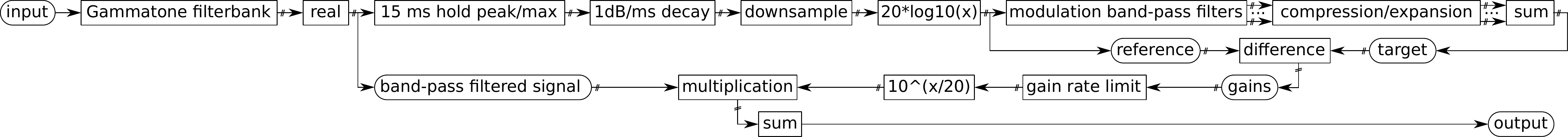}}
	\caption{Diagram illustrating the relations of the main signal processing blocks which were used to implement the signal analysis, manipulation, and re-synthesis with PLATT.}
	\label{fig:5}
\end{figure*}

Figure~\ref{fig:5} illustrates the relations between the signal processing blocks that were used to implement this functionality.
A detailed description is provided in the following.
The exact implementation details are provided in a reference implementation that is written in C (cf. Section~\nameref{sec:ressources}).

\begin{figure}
	\centerline{\includegraphics[width=.85\columnwidth]{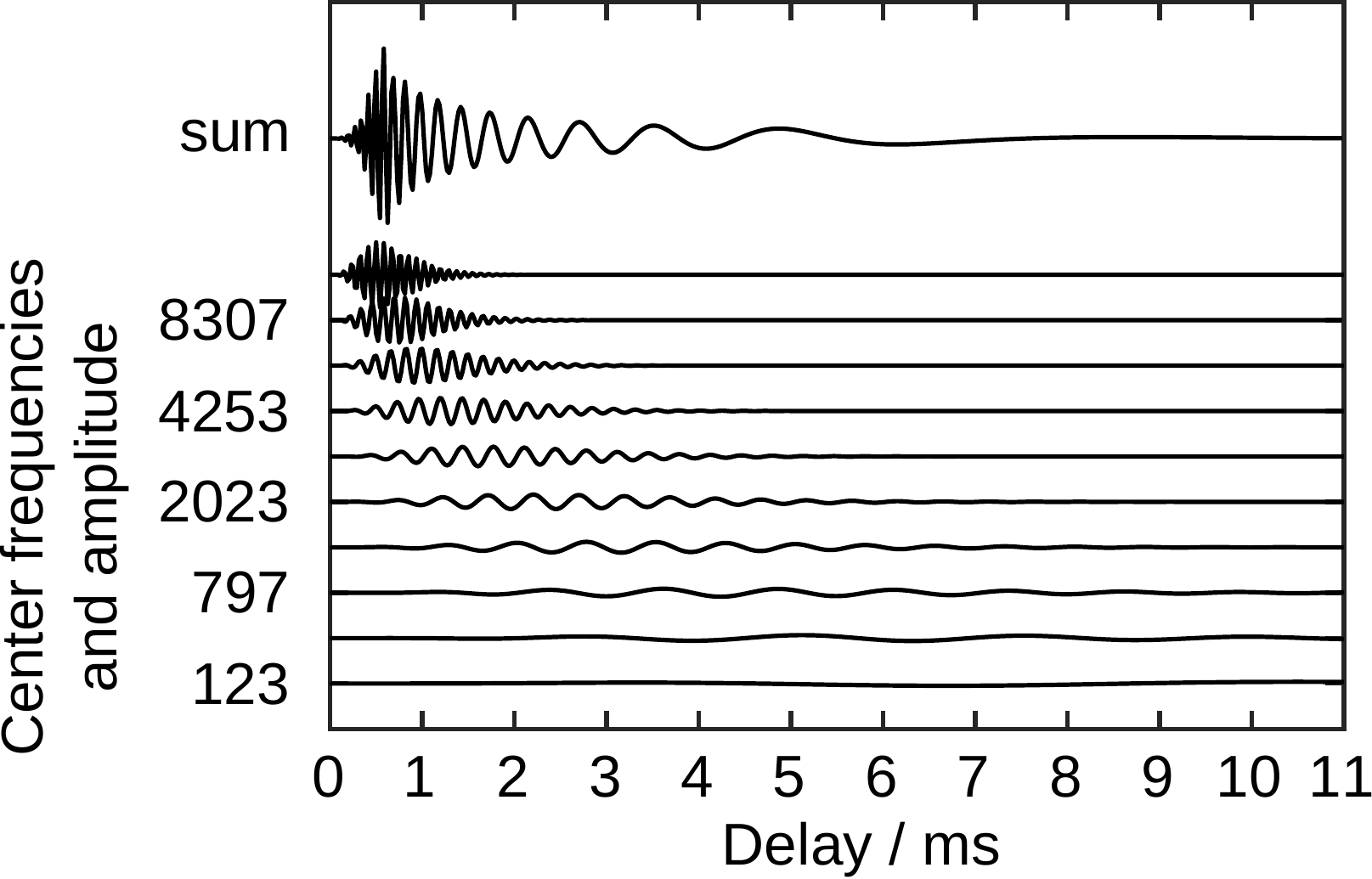}}
	\caption{Real part of the impulse responses of a subset of the normalized, phase-adjusted, fourth-order Gammatone filters that were employed for the frequency decomposition, and the (scaled) sum of the impulse responses of all employed Gammatone filters.}
	\label{fig:6}
\end{figure}

\begin{figure}
	\centerline{\includegraphics[width=.65\columnwidth]{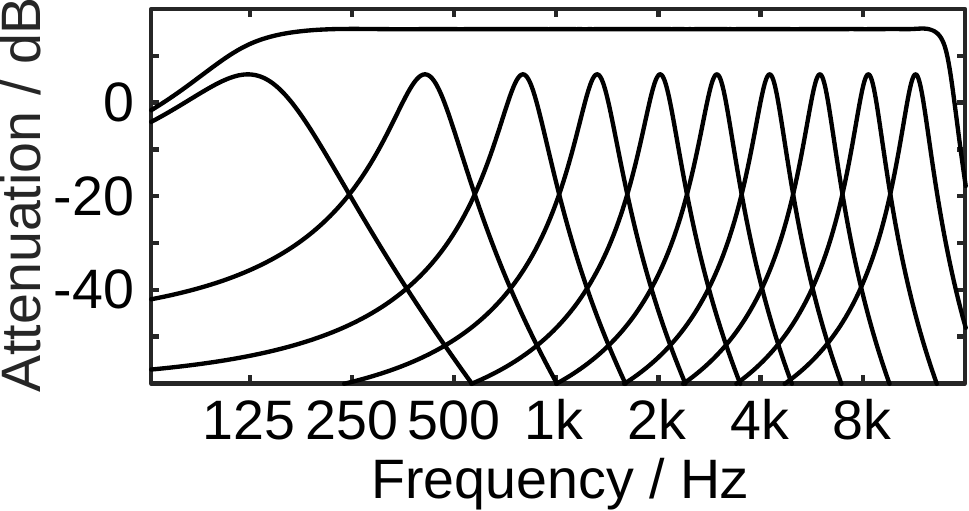}}
	\caption{Absolute values of the transfer functions corresponding to the impulse responses shown in Figure~\ref{fig:6} and the absolute value of the joint transfer function of all filters (including those not shown).}
	\label{fig:7}
\end{figure}

\paragraph{Frequency decomposition \& re-synthesis}
The frequency decomposition of the input signal is performed with a filter bank of fourth-order Gammatone filters.
To get a set of frequencies which are relevant for (automatic) speech recognition, the center frequencies are chosen equidistantly on a Mel-frequency scale with half the distance that is commonly used for calculating MFCCs.
Considering frequencies in the range from $64\,\text{Hz}$ to $16000\,\text{Hz}$ results in the following $78+4$ values: \begin{footnotesize}(64, 93,) 123, 155, 187, 221, 256, 293, 330, 370, 410, 453, 496, 542, 589, 638, 689, 742, 797, 854, 914, 975, 1039, 1105, 1174, 1245, 1319, 1396, 1476, 1559, 1645, 1734, 1827, 1923, 2023, 2127, 2235, 2346, 2462, 2583, 2708, 2838, 2972, 3112, 3257, 3408, 3565, 3727, 3896, 4071, 4253, 4441, 4637, 4840, 5051, 5270, 5498, 5734, 5979, 6233, 6497, 6771, 7056, 7352, 7658, 7977, 8307, 8650, 9006, 9376, 9760, 10158, 10572, 11001, 11447, 11909, 12390, 12888, 13406, 13943, (14501, 15080)\,Hz\end{footnotesize}.
Only the 78 filters with center frequencies from 123\,Hz to 13943\,Hz, spaced approximately 0.5 equivalent rectangular bandwidths (ERB), are used.
The -10\,dB-bandwidth of each fourth-order Gammatone filter is chosen to be equal to the difference of the frequencies two positions right and left to its center frequency, e.g., $221\,\text{Hz}-93\,\text{Hz}=128\,\text{Hz}$ for 155\,Hz.
The aim is to evenly cover the relevant frequency range with filters that have a bandwidth similar to auditory filters ($\approx1\,\text{ERB}$) and allow a trivial re-synthesis in the time domain by simple summation of all filter bank outputs.
With this goal, the filter coefficients are determined as follows:
The pole in the complex z-plane that describes the frequency-dependent properties of a first-order infinite impulse response Gammatone filter is calculated according to the formula
\begin{equation}
p = \left(1-\frac{1}{\sqrt{2} \cdot \frac{10000}{\text{bw}} + 0.5}\right) \cdot \exp\left(2\pi \text{i} \frac{\text{f}_\text{c}}{\text{f}_\text{s}}\right),
\end{equation}
where bw is the -10\,dB-bandwidth in Hz, $\text{f}_\text{c}$ the center frequency of the corresponding fourth-order filter, and $\text{f}_\text{s}$ the sampling frequency.
The phase of the single FIR coefficient of each filter is chosen such that the phases of each pair of fourth-order filters with neighboring center frequencies were identical at the delay where the product of their respective temporal envelopes reaches its maximum.
This happens to be the case at different delays for different pairs and minimizes destructive interference between the corresponding filter outputs.
In addition, the absolute value of the FIR coefficient of each filter is chosen such that the maximum gain is 2.
Together, a) evenly covering frequencies, b) avoiding destructive interference, and c) normalizing the maximum gain result in a very flat frequency response for the sum of all filter bank channels.
Figure~\ref{fig:6} shows the first 11\,ms of the real part of the impulse responses of a subset of the Gammatone filters and also the (scaled) sum over the real parts of all impulse responses.
The joint impulse response (sum of the real-valued impulse responses of all normalized, phase-adjusted, forth-order Gammatone filters) is a downward frequency-sweep.
The frequency-dependent delay can be read from Figure~\ref{fig:6} and is about $2.5\,\text{ms}$ at $2\,\text{kHz}$ and about $4.5\,\text{ms}$ at $800\,\text{Hz}$.
Figure~\ref{fig:7} shows the corresponding absolute values of the transfer functions of the same sub-set of filters and the absolute values of the transfer function corresponding to the joint impulse response of all filters.
The joint transfer function of all 78 employed Gammatone-filters, which characterizes the system property after re-synthesis if the amplitudes are not manipulated, has a flat frequency response between about $150\,\text{Hz}$ and $13\,\text{kHz}$.
That the frequency decomposition and re-synthesis has almost no spectral, and only a limited temporal effect on processed signals indicates that a perceptually mostly transparent re-synthesis can probably be achieved.
The amplitudes of the filter bank output represent perceptually relevant properties of the signal and can be interpreted as a proxy for the displacement of oscillatory systems with properties similar to those of the respective Gammatone filters, e.g., the human basilar membrane.
The filter bank output can be manipulated directly, e.g., multiplied with a time- and frequency-dependent gain function, prior to the re-synthesis.
The rate of change of the gain functions is limited to 24\,dB per period of the corresponding center frequency to limit channel crosstalk.

\begin{figure*}
	\centerline{\includegraphics[width=1.0\textwidth]{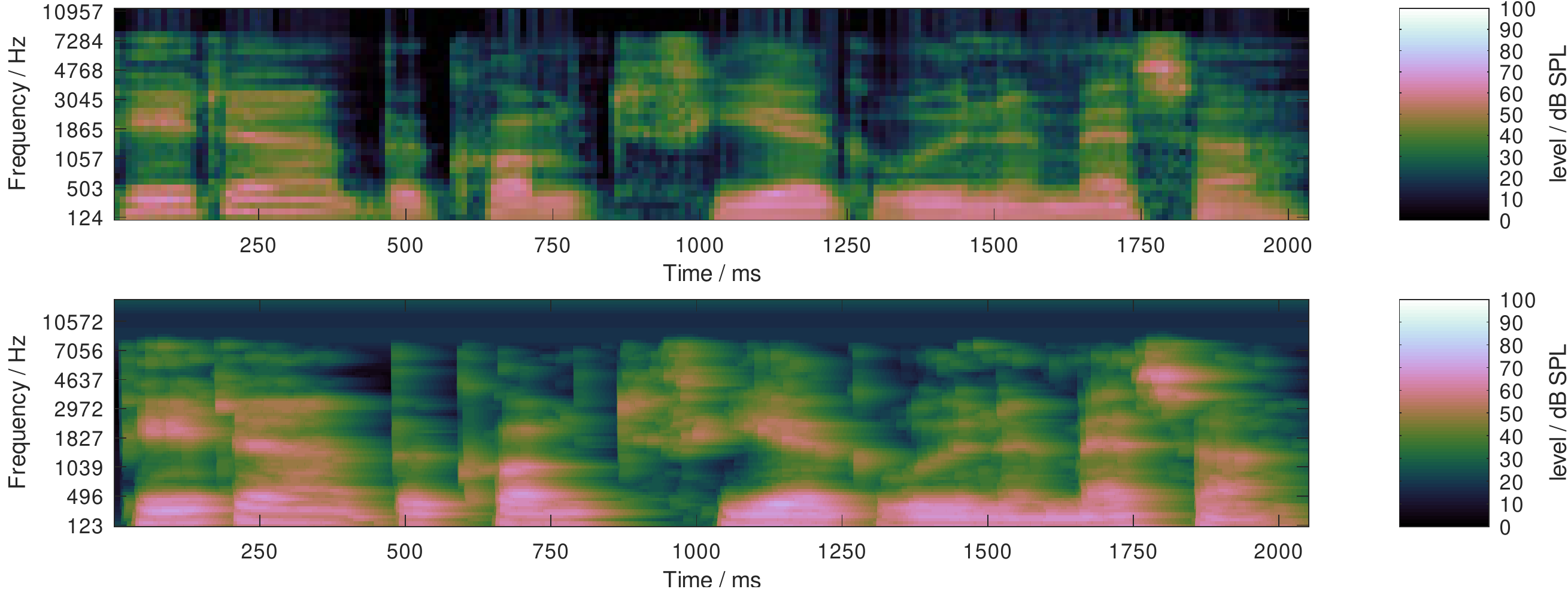}}
	\caption{Comparison of spectro-temporal representations: In the upper panel a log Mel-spectrogram as it is used in FADE of a clean speech sample at 65dB SPL, and in the lower panel the presented specto-temporal representation that is used in PLATT of the same sentence.}
	\label{fig:8}
\end{figure*}

\paragraph{Spectro-temporal signal representation}
A representation of the spectral dynamic which encodes information similar to a log Mel-spectrogram is determined as follows, based on the real-valued output of the filter bank.
For each filter output, the values are held for 15\,ms and decay subsequently with a rate of $1\frac{\text{dB}}{\text{ms}}$, if the held value is above the current value.
This approach approximately extracts the temporal envelope of each channel while preserving fast increases in amplitude (on-sets).
Hence, the exact timing (or temporal fine structure) is removed from this representation, and only the local maximum values remain as an estimate of the maximum amplitude (or displacement) of an oscillatory system with properties similar to those of the employed Gammatone filters.
This representation can be down-sampled by any factor which reduces the sample rate to $\frac{1}{15\,\text{ms}}\approx67\,\text{Hz}$ or higher, without missing any local maximum value.
The encoded information is very similar to the information encoded in features for ASR, where updated spectral values are determined every $10\,\text{ms}$ in frequency-bands which are equally spaced on a Mel-scale.
Unpublished pilot experiments by the author confirm that the proposed representation, down-sampled to $100\,\text{Hz}$, achieves very similar simulation results in a range of speech recognition experiments with FADE.
The use in a hearing device, however, requires faster updates which is why the representation is down-sampled to $1000\,\text{Hz}$, that is, an update of the 78 spectral values is calculated every $1\,\text{ms}$.
In Figure~\ref{fig:8}, the spectro-temporal representation used in PLATT and the log Mel-spectrogram of a clean speech sample at 65\,dB SPL are shown.
Compared to the log Mel-spectrogram, the proposed spectro-temporal representation has a 10-times higher temporal resolution (visible at the on-sets), where, however, the temporal fine structure is effectively removed.
The off-sets are not as prominent because the maximum tracking only allows a decrease by $1\frac{\text{dB}}{\text{ms}} = \frac{\text{100\,dB}}{\text{100\,ms}}$.
Also, the proposed spectro-temporal representation has approximately twice the number of frequency channels, 78 compared to 36 with the the log Mel-spectrogram.
The aim of the spectral oversampling is to provide headroom for spectral modulation manipulations.

Because a pure tone is not changed in amplitude by a filter of the employed Gammatone filter bank when it matches the filters center frequency, the calibration of its input and output values is identical.
Assuming a calibrated setup, only values above the frequency-dependent normal hearing threshold, as defined by the International Organization for Standardization in its standard 226:2003 \citep{iso2003}, are considered in the representation; values below the normal hearing threshold are replaced by the corresponding threshold value.
This effectively removes spectro-temporal modulations that cannot be perceived by listeners with normal hearing from the proposed spectro-temporal signal representation.
The effect can be observed in the high frequency range ($>8000$\,Hz) in Figure~\ref{fig:8}, where the speech signal has no energy.

\paragraph{Adaptive spectral gain}
The adaptive determination of time- and frequency-dependent gains takes into account the current spectral input dynamic and the currently available output dynamic.
It aims to minimize the compression with the constraint to avoid masking the signal parts which could carry important (speech) information.
It also allows to expand the spectral modulations that are assumed to be important for speech recognition and trade the such increased signal dynamic against an increased compression of less relevant signal dynamic.

\begin{figure*}
	\centerline{\includegraphics[width=.80\textwidth]{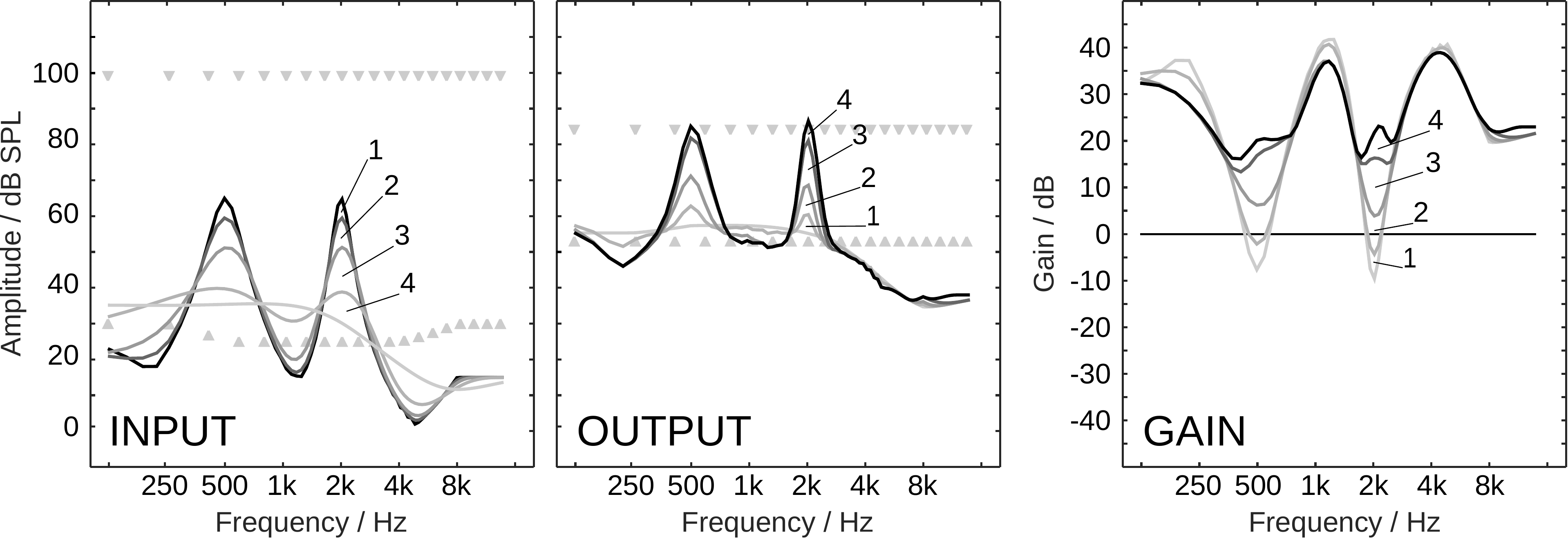}}
	\caption{Example dynamic mapping of two pure tones. Left panel: Input dynamic analysis with spectral modulation low-pass filters. Center panel: (Conditional) reconstruction with reduced dynamic. Right Panel: Gains required to map the input dynamic to the reconstruction stages of the output dynamic (final gains are black). Numbers indicate the differences.}
	\label{fig:9}
\end{figure*}

The spectral input dynamic is analyzed with spectral modulation low-pass filters.
For this, each vector of the $78$ spectral values is convolved with Hanning windows of the following widths to obtain increasingly spectrally smoothed versions of the initial vector: $8$, $16$, $32$, and $64$, which approximately correspond to a full width at half maximum (FWHM) of $2$, $4$, $8$, and $16\,\text{ERB}$, respectively.
The left panel of Figure~\ref{fig:9} shows an example of the spectral analysis for a signal which consists of two pure tones, of $500\,\text{Hz}$ and $2000\,\text{Hz}$.
The input spectral values are depicted in black, the smoothed versions with ascending widths of the Hanning window in increasingly lighter shades of gray.
The differences between the curves of increasingly smoothed spectral representations are indicated with the digits 1 to 4.
Because the difference of two low-pass filters is a band-pass filter, the differences 1, 2, 3, and 4 can be interpreted as the result of a spectral modulation band-pass filtering which roughly contains the respective spectral modulation frequencies 1) above $\frac{1}{4}$, 2) from $\frac{1}{4}$ to $\frac{1}{8}$, 3) from $\frac{1}{8}$ to $\frac{1}{16}$, and 4) from $\frac{1}{16}$ to $\frac{1}{32}$\,$\frac{\text{cycles}}{\text{ERB}}$, respectively.
By definition, the original vector of spectral values (cf. black line in left panel of Figure~\ref{fig:9}) can be recovered by adding the differences 1 to 4 to the low-pass filtered spectral values which contain the spectral modulation frequencies below $\frac{1}{16}$\,$\frac{\text{cycles}}{\text{ERB}}$ (cf. lightest line in left panel of Figure~\ref{fig:9}).
In the following, the vector of low-pass filtered spectral values which contains the lowest modulation frequencies is referred to as the \emph{base layer}.
The base layer contains a very coarse description of the spectral dynamic and resolves spectral patterns larger than approximately $16\,\text{ERB}$.
The other layers, the differences 4 to 1, resolve the spectral dynamic which is not described in the base layer with decreasing pattern sizes.

The base layer needs to be mapped from the input dynamic range to the output dynamic range.
For the application in a hearing device, the mapping of the base layer could be performed with a prescription rule.
In addition, the limits of the input and output dynamic ranges need to be defined.
Here, for the input, a dynamic range that is probably relevant for normal hearing listeners is assumed: A frequency-independent uncomfortable level of $105\,\text{dB~SPL}$ as the upper limit, and levels of $25\,\text{dB~SPL}$ from $500\,\text{Hz}$ to $4\,\text{kHz}$, and $30\,\text{dB}$ above $8\,\text{kHz}$ and below $250\,\text{Hz}$, as the lower limit.
The assumed input dynamic range is indicated with triangles in the left panel in Figure~\ref{fig:9}.
For the output, an exemplary reduced dynamic range is assumed, which is arbitrarily limited to frequency-independent levels from $50$ to $90\,\text{dB~SPL}$; resembling elevated hearing thresholds due to environmental noise or impaired hearing and a lower acceptance of high levels.
The targeted output dynamic range is indicated with triangles in the center panel in Figure~\ref{fig:9}.
The mapping of the base layer can be independent from the input and output dynamic range definitions.
In the context of a hearing device, the mapping of the base layer will mainly determine the output levels and strongly affect loudness perception, while the lower limit of the output dynamic range (to be chosen related to the individual hearing thresholds) will strongly affect speech recognition performance.
The defined output dynamic range is the reservoir that can be used by PLATT to map the input dynamic.

In our example in Figure~\ref{fig:9}, let's assume the base layer was mapped linearly from the defined input dynamic range to the defined output dynamic range.
The base layer is depicted as the lightest gray line in the left panel, and the mapped base layer as the lightest gray line in the center panel.
The gains which would be theoretically required to achieve such a smooth output dynamic, given the black line in the left panel as the input dynamic, are depicted as the lightest gray line in the right panel.
But such an extreme compression of the spectral dynamic would introduce audible artifacts and would most likely have a negative effect on speech recognition performance.
Hence, the more of the remaining original signal dynamic described by the differences 1 to 4 is added back to the mapped base layer, the less compression will be needed, and the better spectral modulation patterns will be preserved.
However, unconditionally adding the whole dynamic that is encoded in the differences 1 to 4 could result in output levels below the lower limit of the output dynamic range, which might not contribute to speech recognition anymore, or above the upper limit of the output dynamic range, which might lead to undesirably high output levels.
A good compromise in the fundamental conflict that too much and too little compression can both result in sub-optimal speech recognition performance requires a compression management which depends on the current spectral input dynamic and the currently available output dynamic.

To prefer spectral patterns which are important for robust (automatic) speech recognition, the differences corresponding to high spectral modulation frequencies are added first.
The highest spectral modulation frequencies are encoded in difference 1 and account only for a very small part of the total dynamic, which is why it is added \emph{unconditionally}.
The example with two pure tones is an extreme one which assesses the maximum dynamic that can be encoded in each difference, which is about 6\,dB for difference 1.
The corresponding output dynamic, when only adding difference 1 to the base layer, can be observed in the center panel of Figure~\ref{fig:9} as the light gray curve which only deviates slightly from the base layer.
Probably the most important spectral modulation frequencies describe spectral patterns between $2$ and $4\,\text{ERB}$ and are mainly encoded in difference 2, which is why it is also added unconditionally.
To protect this difference, which encodes a maximum dynamic of less than 9\,dB, against a hearing loss of class D as implemented in FADE with the level uncertainty, it can be expanded by a factor greater than 1 prior to adding it to the base layer.
The expansion of the difference 2 could increase the total output signal dynamic, which however can often be compensated by an increased compression of the remaining differences 3 and 4, which encode larger part of the signal dynamic, compared to difference 2.

The remaining differences 3, and subsequently 4, are added \emph{conditionally} and possibly partially according to the following constraints:
1) For each frequency, the respective difference is multiplied with the highest possible factor between 0 and 1 for which its addition to the current output dynamic (which already includes the base layer and differences 1 and 2) does not result in output values above the upper limit or below the lower limit.
For example, if the current value is $80\,\text{dB~SPL}$, the difference $+30\,\text{dB~SPL}$, and the limit $90\,\text{dB~SPL}$ only a third of difference can be added and the factor is $\frac{1}{3}$.
This rule is not sufficient because, when applied independently for each frequency band, it would result in a total loss of high spectral modulation frequencies in the frequency regions where no output dynamic range is left to add the difference; the already added spectral modulations needs to be protected from being compressed.
Hence, 2) the minimum factor (highest compression) is propagated along the frequency axis by adding values of normalized Hanning windows of FWHM of 6 and 12 samples for difference 3 and 4, respectively.
For example, if the factor according to 1) turns out to be zero at the center frequency of $2708\,\text{Hz}$ when adding difference 3, the factor at $2346\,\text{Hz}$ and $3112\,\text{Hz}$ can be at most $0.5$ and below $2023\,\text{Hz}$ and above $3565\,\text{Hz}$ again 1.
The propagation ensures that neighboring channels are compressed similarly and protects any higher spectral modulation frequencies.
The final desired spectral output levels are described by the sum of the mapped base layer, the unconditionally added differences 1 and 2, and the conditionally compressed differences 3 and 4.
The frequency-dependent gain needed to achieve the desired spectral output levels is the difference of the spectral output levels (black curve in the center panel in Figure~\ref{fig:9}) and spectral input levels (black curve in the left panel in Figure~\ref{fig:9}).
The final frequency-dependent gain is plotted in black in the right panel of Figure~\ref{fig:9} along with the partial gains that would theoretically be needed after the cumulative addition of only differences 1 to 3 to the base layer in increasingly darker gray shades.
The final frequency-dependent gain is then applied to the output of the Gammatone filter bank with the limitation of the rate of change to 24\,dB per period of the corresponding center-frequencies.

Admittedly, in our example with the two pure tones, there is only energy at 500\,Hz and 2000\,Hz, and hence only the level of the two tones will be changed, while the gains at other frequencies will have no effect.
However, this signal creates a pattern of extreme spectral modulation in the proposed signal analysis, and hence is well suited to illustrate how PLATT works.
For a signal with only low spectral modulations, no conditional compression would be required.

\begin{figure*}
	\centerline{\includegraphics[width=.8\textwidth]{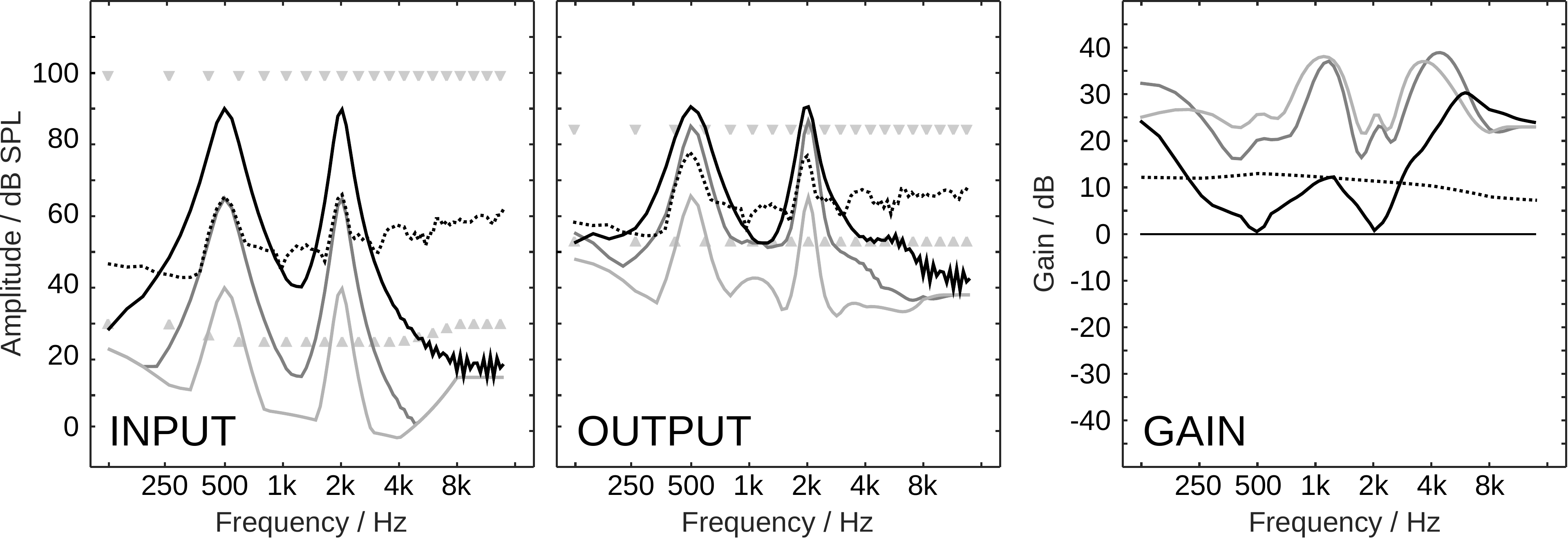}}
	\caption{Examples of calculated spectral gain for an example with reduced available output dynamic range as indicated by the triangles.
	Left panel: Spectral input dynamic with two pure tones at different presentation levels (gray shaded solid curves), and with white noise added (dotted curve).
	Center panel: Corresponding spectral output levels.
	Right panel: Corresponding gains.}
	\label{fig:10}
\end{figure*}

\paragraph{Summary of PLATT dynamic range manipulation}
With PLATT, compression is only applied if the available output dynamic range is less than required to represent the input signal dynamic.
The main effect can be observed in Figure~\ref{fig:10}, where the calculated gains for \emph{high} (solid curves) and \emph{reduced} (dotted curves) spectral input dynamic are shown for an exemplarily reduced output dynamic range, as indicated by the triangles.
While the signals with high spectral dynamic are compressed (observed here as different gains for different frequencies) to make relevant signal portions audible and avoid excessively high output levels, the signal with low spectral dynamic is not (observed here as similar gains for different frequencies), because it fits in the available output dynamic range without compression.
Only low spectral modulation frequencies, which are less important for speech recognition, are conditionally compressed, while the important higher spectral modulation frequencies can be even expanded.
An efficient implementation of spectral modulation filtering in the time-domain is possible with reasonably low latency, fast reaction times, and probably little audible artifacts due to the auditory-motivated signal decomposition and re-synthesis.

\paragraph{Compensation of level uncertainty with PLATT}
The possibility to selectively expand the spectral modulation patterns between 2 and 4\,ERB with PLATT was used to protect these patterns against the effect of the level uncertainty as implemented in FADE.
With the aim to best decouple the evaluation of the expansion from the conditional compression feature of PLATT, the available output dynamic range was set to match the input dynamic range, and the base layer mapping was set to identity.
This minimizes compression and effectively disables the fine-grained compression management for all but very low and very high signal levels.
Nonetheless, in view of expanding a part of the signal dynamic, such a management will probably be needed in an application with a limited output dynamic range to counter the expansion with a possible compression of other parts of the signal dynamic.

\begin{figure*}
	\centerline{\includegraphics[width=1\textwidth]{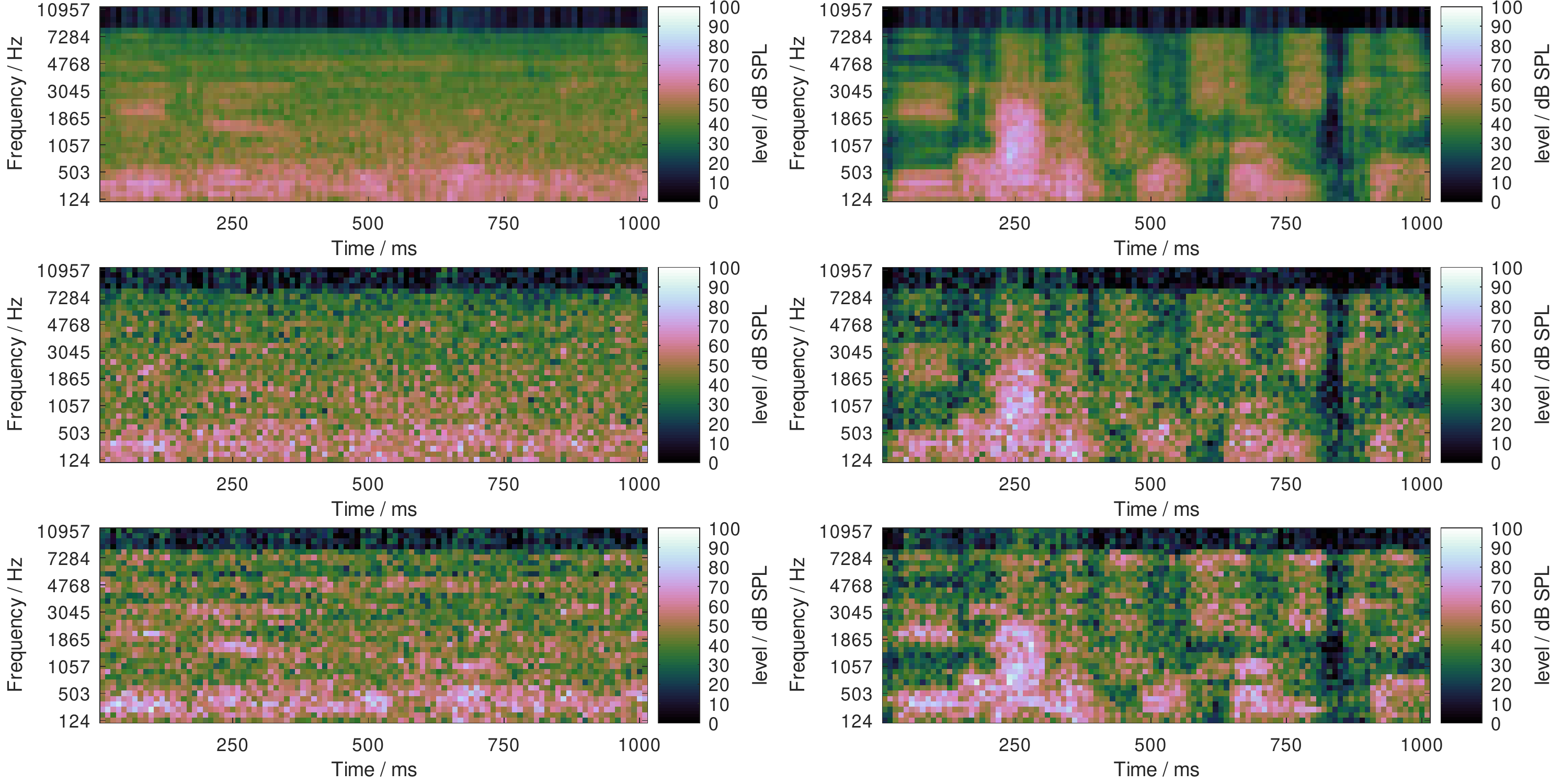}}
	\caption{Illustration of the effect of the dynamic range manipulation with PLATT-6: Log Mel-spectrograms of noisy speech in stationary (left column) and fluctuating noise (right column) at 0\,dB SNR (top row), the same log Mel-spectrograms with level uncertainty of 7\,dB (center row), and log Mel-spectrograms with level uncertainty of 7\,dB of the same signals processed with PLATT-6, that is with an expansion factor of 6.}
	\label{fig:11}
\end{figure*}

Expansion factors of 2, 4, 6, and 8 were considered for the evaluation, and the corresponding compensation conditions are referred to as PLATT-2 to PLATT-8 in the remainder of the manuscript.
The effect of processing noisy speech signals at 0\,dB SNR with PLATT-6 is illustrated in Figure~\ref{fig:11}.
In the top row, log Mel-spectrograms of a speech signal in the stationary and fluctuating noise at 0\,dB SNR are depicted in the left and right panel, respectively.
In the center row, the same log Mel-spectrograms are shown, but with a level uncertainty of 7\,dB.
In the bottom row, log Mel-spectrograms of the same signals, but processed with PLATT-6 are shown with a level uncertainty of 7\,dB.
Especially for the stationary noise, some of the speech patterns are better distinguishable in the log Mel-spectrogram with the added level uncertainty \emph{after} the processing.
This justifies the expectation that the expansion might protect important speech patterns against level uncertainty.
With the fluctuating masker, the speech patterns are not as easily distinguishable from the background noise.
But comparing the representation of the unprocessed (center) and processed (bottom) signals, more of the patterns that are hidden by the level uncertainty in the unprocessed signal are discernible in the processed variant.
This illustrates that the expansion does not selectively process the speech portions of the signal but generally protects the spectral modulation patterns between 2 and 4\,ERB against the level uncertainty, independently of what caused these patterns.
Also observable, the processing with PLATT-6 results in a possible increase in the output level.
This affirms the need to evaluate the method in a context where simple time-invariant linear amplification will not change the speech recognition performance and thus decouple the effect of the expansion from the effect of any linear amplification.

The just described effects of an expansion of the spectral modulation patterns between 2 and 4\,ERB on noisy speech signals were quantified in simulation experiments with FADE.
For this, SRTs with the German matrix sentence test were simulated for all combinations of considered noise maskers (OLNOISE, ICRA5-250), noise presentation levels (0, 10, 20, ..., 100\,dB SPL), listener profiles (P-8000-1 through P-1000-21), and compensations (none, and PLATT-2 through PLATT-8); summing a total of 1760 outcome predictions with FADE.

\subsection*{Evaluation of simulation results}
Key simulation outcomes are presented as \enquote{Plomp curves}, that is, SRT-50s in dB~SPL as a function of the noise presentation level, analog to Figure~\ref{fig:2}.
On the one hand, this depiction allows to assess the effect of the (noise) presentation level on the predicted SRTs.
On the other hand, it also allows to quantify improvements in SRT in aided conditions over a given reference condition, e.g., normal hearing or an unaided condition.
Because not all 1760 data points can presented in graphical form in this contribution, a summary of the achieved improvements in SRT at high presentation levels, at which we can confidently assume that linear time-invariant amplification cannot improve the SRT, are presented in the form of a table for all listener profiles and compensations.
For key listener profiles, psychometric functions were obtained by simulating SRT-20 to SRT-90 to assess the SNR-dependency of PLATT.
The main interest here was if the effect of the PLATT compensation is different for higher SRTs, which are more realistic for conversations.

\subsection*{Availability of resources}
\label{sec:ressources}
To facilitate the reproduction of the presented experiments, and to encourage, foster, and accelerate the verification and adoption of the presented methods, the employed resources are provided, as far as licensing allows it.
FADE\footnote{\url{https://doi.org/10.5281/zenodo.4003779}} version 2.4.0, which is open source sofware, was used for the FADE simulations.
The code and scripts for the setting up the simulations were based on, and are now integrated into the measurement and prediction framework\footnote{\url{https://doi.org/10.5281/zenodo.4500810}}.
This includes:
\begin{itemize}
	\item The modified feature extraction.
	\item The reference implementation of PLATT and the used configuration files
	\item The scripts which prepare and run the FADE simulations using the modified feature extraction and the reference implementation of PLATT.
	\item The scripts which evaluate raw the experimental results and plot the results figures.
\end{itemize}

This does \emph{not} include:
\begin{itemize}
	\item The Hidden Markov Toolkit\footnote{\url{http://htk.eng.cam.ac.uk/}} (used by FADE), which cannot be distributed because of the license.
	\item The speech material of the German matrix sentence test\footnote{\url{https://www.hoertech.de/en/devices/intma.html}} and the corresponding test specific noise signal (OLNOISE), which cannot be distributed because of the license.
	\item The ICRA5-250\footnote{\url{http://medi.uni-oldenburg.de/download/ICRA/index.html}} noise signal, due to the unknown license conditions.
\end{itemize}

\section*{Results}
\label{sec:results}
\subsection*{Effect of limit frequency and level uncertainty}
\begin{figure*}[h!]
	\centerline{\includegraphics[width=\textwidth]{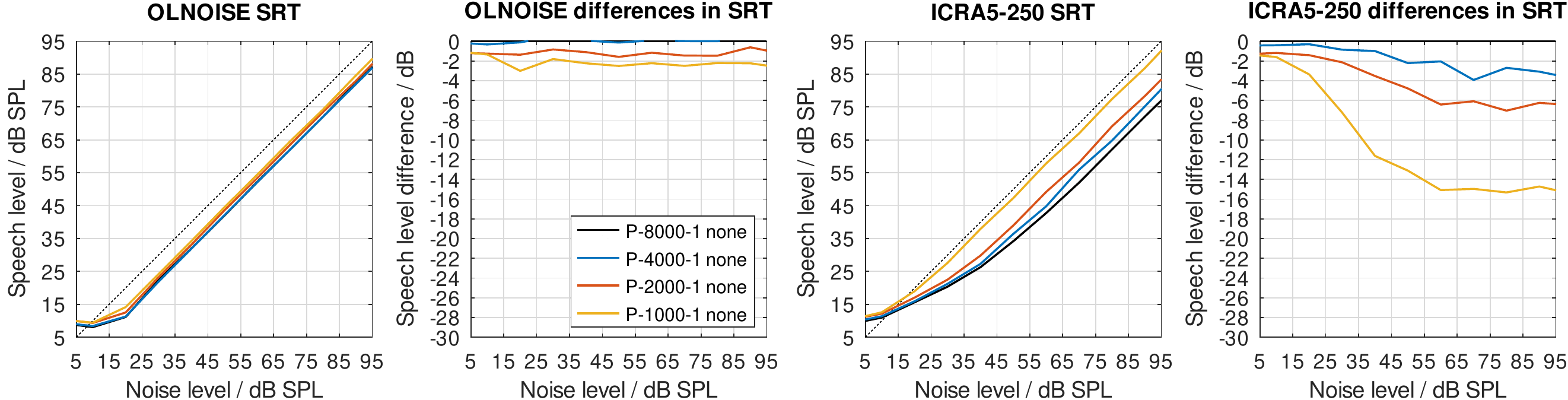}}
	\caption{Simulated \enquote{Plomp curves} in stationary noise (first panel) and fluctuating noise (third panel) when limiting the available frequency range (to 8000, 4000, 2000, and 1000\,Hz) in the feature extraction stage.
	The dotted lines indicate an SNR of 0\,dB.
	The corresponding level-dependent differences in SRT compared the profile plotted in black (here profile P-8000-1) are depicted in panels two and four.
	}
	\label{fig:12}
\end{figure*}

\begin{figure*}[h!]
	\centerline{\includegraphics[width=\textwidth]{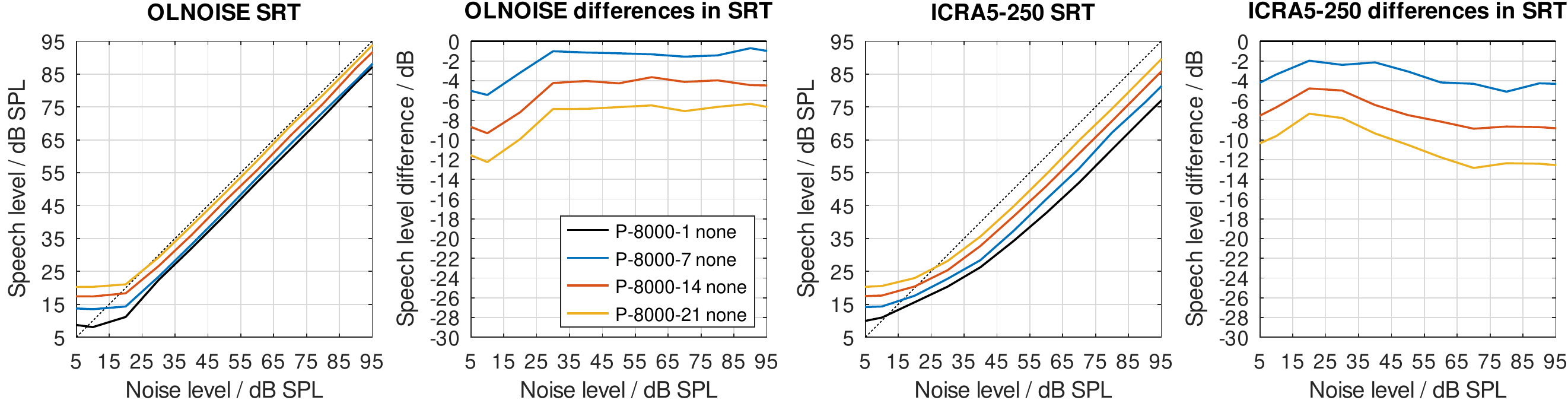}}
	\caption{Effect of increasing the level uncertainty (to 7, 14, and 21\,dB). Analog to Figure~\ref{fig:12}.}
	\label{fig:13}
\end{figure*}

Two modifications were used to implement a hearing loss of class D, the limitation of the frequency range up to a limit frequency, and the increase of the level uncertainty.
Their separate effect on simulated SRTs is shown in Figure~\ref{fig:12} and Figure~\ref{fig:13}, respectively.
\begin{figure*}[h!]
	\centerline{\includegraphics[width=\textwidth]{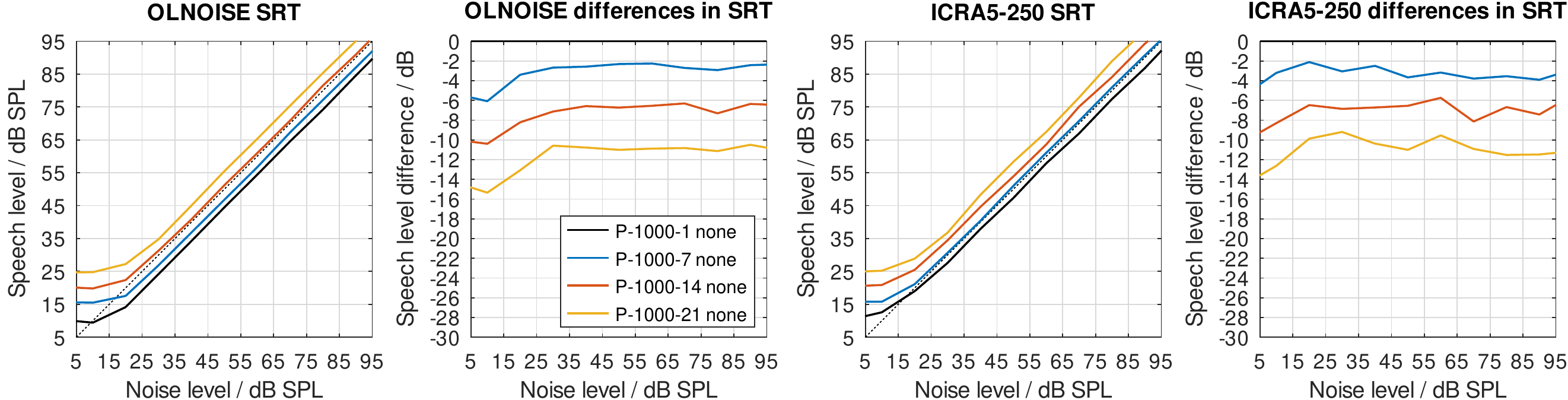}}
	\caption{Effect of increasing the level uncertainty when the frequency range is limited to 1000\,Hz. Analog to Figure~\ref{fig:12}.}
	\label{fig:14}
\end{figure*}
\begin{figure*}[h!]
	\centerline{\includegraphics[width=\textwidth]{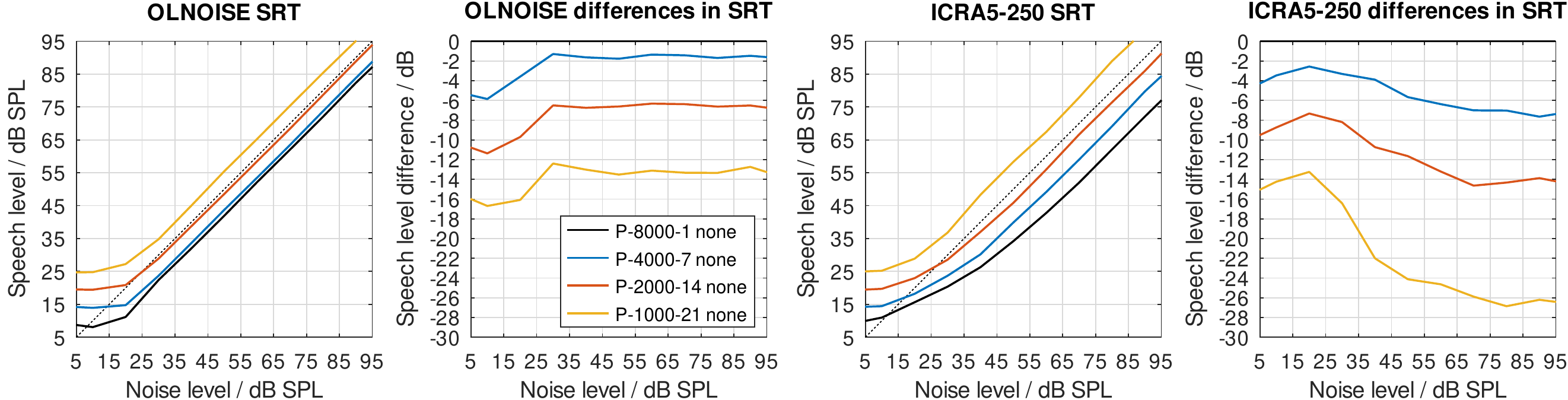}}
	\caption{Effect of mixed profiles with increasing level uncertainty and frequency range limitation. Analog to Figure~\ref{fig:12}.}
	\label{fig:15}
\end{figure*}

The limitation of the available frequency range affected the simulated SRTs in the fluctuating noise condition much more than in the stationary noise condition when assuming a level uncertainty of 1\,dB (cf. Figure~\ref{fig:12}).
In the stationary noise condition, the limitation of the frequency range to 4000\,Hz did not have an effect on the simulated outcome of the German matrix sentence test.
A reduction to 2000\,Hz resulted in a small, mostly level-independent increase in SRT of about 1\,dB, and a further reduction of the frequency range to 1000\,Hz, resulted in a further increase of about 1\,dB.
Hence, limiting the frequency range while assuming a level uncertainty of 1\,dB had relatively little effect on the simulated SRTs.
A different picture can be observed in the fluctuating noise condition.
There, the reduction of the frequency range had a large, and level-dependent effect on the simulated SRTs.
The effect of the reduction increases with the higher presentation level and stabilizes at levels above 60\,dB SPL.
At these levels, the increase in SRT was about 3, 6, and 15\,dB for a limitation to 4000, 2000, 1000\,Hz, respectively.

The increase of the level uncertainty also affected the simulated SRTs in the fluctuating noise conditions more than in the stationary noise condition (cf. Figure~\ref{fig:13}).
The effect was not as level-independent at low presentation levels as one might have expect.
This indicates an interaction between the implementation of the absolute hearing threshold and the level uncertainty of about 5\,dB.
The behavior could already be observed in the data from \cite{kollmeier2016} (cf. right panel in Figure~\ref{fig:2}).
However, in this contribution, the focus does not lie on the interactions of the level uncertainty with the absolute hearing threshold but effective elimination of the absolute hearing threshold as a factor from the evaluation.
In the fluctuating noise condition, this is the case for presentation levels of 70\,dB SPL and above, where the differences (cf. panel four in Figure~\ref{fig:13}) stabilize.
At these levels, the increase in SRT was about 4, 9, and 12\,dB for level uncertainties of 7, 14, and 21\,dB, respectively.
In the stationary noise condition, the corresponding increases were lower, with about 1, 4, and 6\,dB for level uncertainties of 7, 14, and 21\,dB, respectively.

Neither implementation alone achieves to increase the SRT to above 0\,dB with the considered parameter values.
In listeners with impaired hearing, we assume that both factors, the level uncertainty as well as the limited frequency range, contribute to the class D loss.
All combinations of both parameters were considered, of which a few are presented in Figure~\ref{fig:14} and \ref{fig:15} to assess their interaction.
In Figure~\ref{fig:14}, the simulation results for increased values of level uncertainty are shown when the frequency range was limited to 1000\,Hz.
Comparing the results to Figure~\ref{fig:13} (limitation to 8000\,Hz), the limitation of the frequency range increased the effect of the level uncertainty in the stationary noise condition, while the effect in the fluctuating noise condition remained similar.
In the stationary noise condition, the average increase in SRT at levels above 70\,dB was about 3, 7, and 11\,dB for level uncertainties of 7, 14, and 21\,dB, respectively.
In the fluctuating noise condition, the corresponding increases were very similar, with 4, 7, and 11\,dB for level uncertainties of 7, 14, and 21\,dB, respectively.

It cannot be expected that a partial expansion of the signal dynamic (e.g. with PLATT) will compensate the increase in SRT due to limiting the frequency range.
However, the simulations suggest an interaction of the level uncertainty with the limited frequency range, more specifically, that an increase in level uncertainty would be worse in the stationary noise condition if the frequency range is limited.
This is due to the strongly non-linear nature of the (automatic) speech recognition process, which can be interpreted to involve a forward error correction scheme that integrates over frequency and over time, and which needs to fail in order to achieve an error rate as high as 50\%.
Forward error correction requires redundancy, which is reduced by limiting the frequency range.
The simulation result shows that considering level uncertainty and a limitation of the frequency range separately would not show the full picture.

To graphically present effect of the PLATT compensation on the simulated SRTs, the mixed profiles P-4000-7, P-2000-14, and P-1000-21 were used.
This choice reflects the assumption that the limitation of the frequency range (due to the audiogram) and an increase in level uncertainty are probably correlated to some extent.
The simulations results with these listener profiles without PLATT compensation are shown in Figure~\ref{fig:15}.
Compared to the normal-hearing configuration, in the stationary noise condition, the average increases in SRT at levels above 70\,dB were about 2, 6, and 13\,dB with the profiles P-4000-7, P-2000-14, and P-1000-21, respectively.
In the fluctuating noise condition, the corresponding increases were much more pronounced, with 7, 14, and 26\,dB with the profiles P-4000-7, P-2000-14, and P-1000-21, respectively.

\subsection*{Effect of PLATT expansion}
\begin{figure*}[h!]
  \centerline{\includegraphics[width=\textwidth]{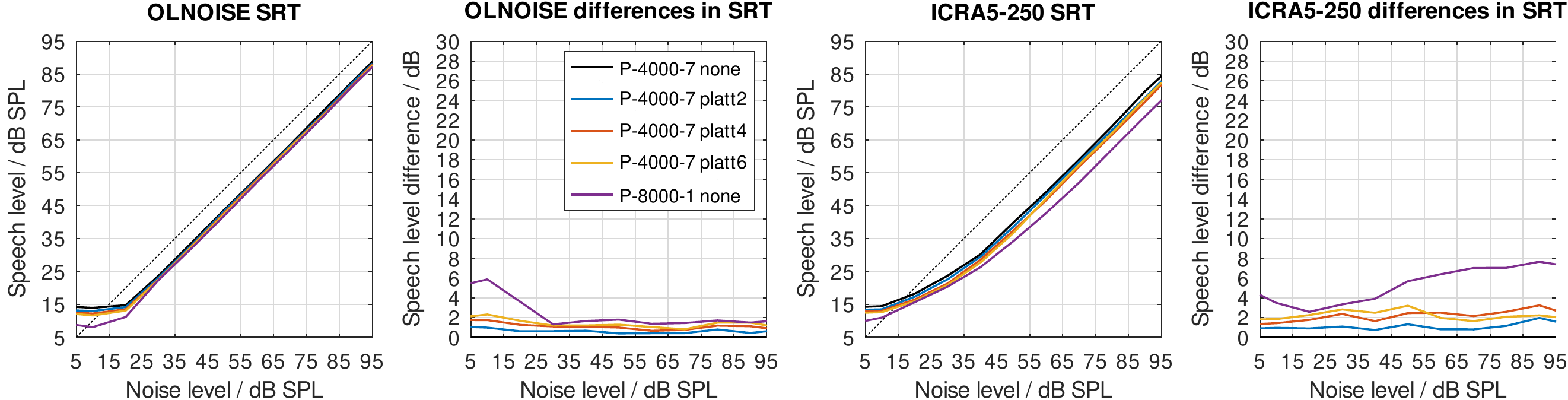}}
	\caption{Simulated \enquote{Plomp curves} for listener profile P-4000-7 without and with PLATT expansion factors 2, 4 and 6 in stationary noise (first panel) and fluctuating noise (third panel).
	The dotted lines indicate an SNR of 0\,dB.
	The corresponding level-dependent differences in SRT compared the profile plotted in black (here the unaided profile P-4000-7) are depicted in panels two and four.
	The data with profile P-8000-1 (normal hearing) was added as an orientation for normal-hearing performance.}
  \label{fig:16}
\end{figure*}

\begin{figure*}[h!]
  \centerline{\includegraphics[width=\textwidth]{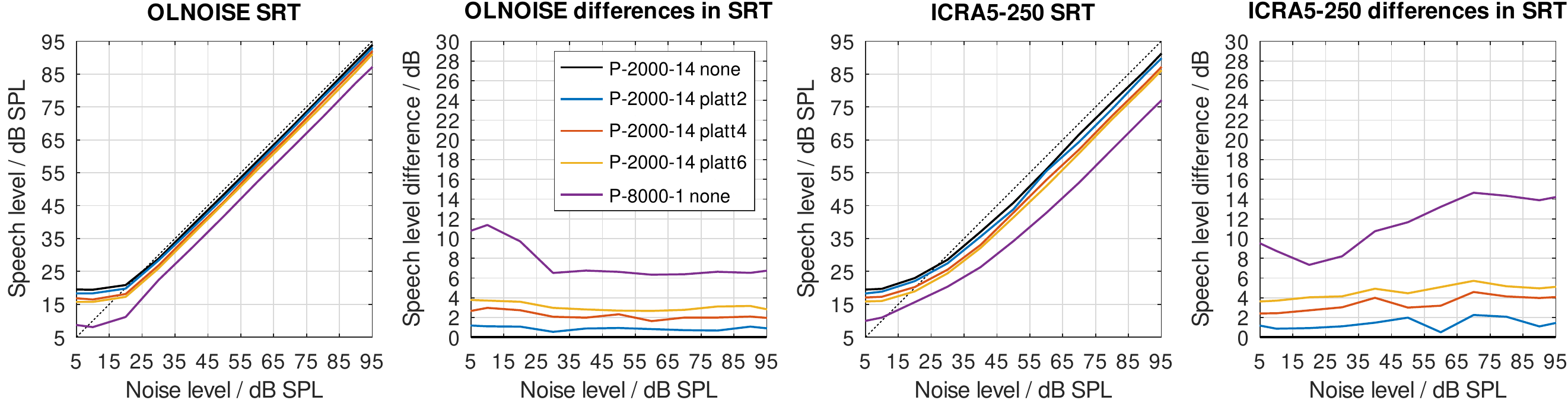}}
  \caption{Simulated \enquote{Plomp curves} for listener profile P-2000-14 without and with PLATT. Analog to Figure~\ref{fig:16}.}
  \label{fig:17}
\end{figure*}

\begin{figure*}[h!]
  \centerline{\includegraphics[width=\textwidth]{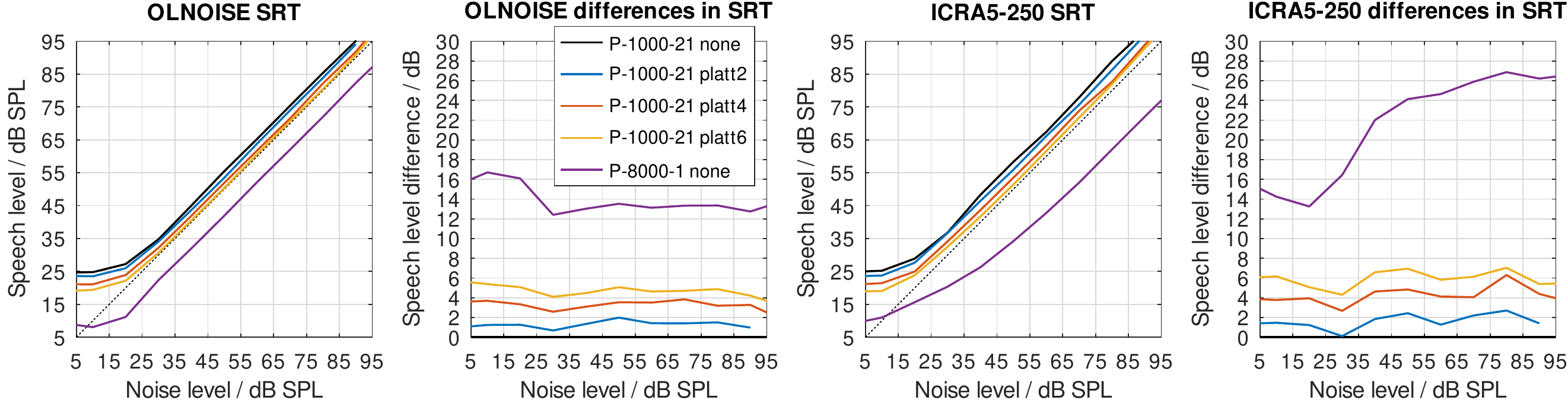}}
  \caption{Simulated \enquote{Plomp curves} for listener profile P-1000-21 without and with PLATT. Analog to Figure~\ref{fig:16}.}
  \label{fig:18}
\end{figure*}

The effect of the expansion with PLATT on the simulation results for listener profiles P-4000-7, P-2000-14, and P-1000-21 is presented in Figures~\ref{fig:16}, \ref{fig:17}, and \ref{fig:18}, respectively.
The blue, red, and yellow lines indicate the simulated speech recognition performance with compensations PLATT-2, PLATT-4, and PLATT-6, respectively.
The black lines indicated the corresponding unaided reference conditions, while the purple lines indicate the simulated performance of profile P-8000-1, that is, the performance of a listener with normal hearing.

The expansion with PLATT improved the SRTs in almost all simulated conditions.
The benefit in SRT due to PLATT in was generally higher in the fluctuating noise condition than in the stationary noise condition.
Higher expansion factors strongly tended to result in higher benefits, however, not in all conditions.
As expected, the observed compensations were partial, that is, the simulated performance of listeners with normal hearing were not restored.
For a quantitative analysis, the results are interpreted only at the levels at which an improvement in SRT due to simple linear amplification can be ruled out.

The curves stabilize at high SRTs, that is, a further increase in levels does not improve the SNR anymore.
In the stationary noise condition, this is the case for presentation levels of 40\,dB SPL and above.
But even above this level, the curves show some variability, which is due to the stochastic nature of the simulation process.
The plotted data points were derived as the difference of two independent simulations.
Assuming a random (normally distributed) prediction error of 0.5\,dB, this would lead to a random error of the plotted data points of approximately $\pm0.71$\,dB, that is for 5 of 100 simulation results, on average, the error will be larger than $\pm1.96\cdot0.71\approx\pm1.39$\,dB.
Hence, \enquote{stabilized} refers to reaching a state which may allow to assume that the remaining variability is due to the random error.
In the fluctuating noise condition, the random error is increased, because of the generally shallower slope of the psychometric function in this condition.
There, the curve of the normal-hearing profile stabilizes at levels of about 70\,dB.
The curves of the unaided and compensated simulations already stabilize at lower levels.
Hence, at noise presentation levels of 70, 80, and 90\,dB SPL an improvement in SRT due to simple linear amplification can be ruled out.

For a break-down of the simulation results, the average improvements at noise presentation levels of 70, 80, and 90\,dB SPL were calculated and are reported in Table~\ref{tab:1}.
\begin{table}[h]
	\small\sf\centering
	\caption{\label{tab:1}
	 Simulated benefits in SRT compared to the respective unaided conditions averaged over high presentation levels (70, 80, and 90\,dB SPL), where simple linear amplification cannot improve the SRT. NH indicates the normal-hearing listener profile.
	}
	\begin{tabular}{l|rrrr|r}
		\multicolumn{6}{r}{Stationary noise (OLNOISE)}\\
		\hline
		\hline
		Profile & \footnotesize \footnotesize PLATT-2 & \footnotesize PLATT-4 & \footnotesize PLATT-6 & \footnotesize PLATT-8 & NH\\
		\hline
		P-8000-1  &  0.0 &  0.1 &  0.2 & -0.1 &  0.0\\
		P-8000-7  &  0.5 &  0.7 &  0.8 &  0.7 &  1.2\\
		P-8000-14 &  1.0 &  2.0 &  2.5 &  2.7 &  4.2\\
		P-8000-21 &  0.8 &  2.2 &  3.2 &  3.9 &  6.7\\
		\hline
		P-4000-1  &  0.1 &  0.4 &  0.4 &  0.2 & -0.1\\
		P-4000-7  &  0.5 &  1.0 &  1.3 &  1.2 &  1.5\\
		P-4000-14 &  0.9 &  2.2 &  3.0 &  3.4 &  4.9\\
		P-4000-21 &  1.3 &  3.0 &  4.0 &  4.8 &  7.9\\
		\hline
		P-2000-1  &  0.1 &  0.2 &  0.2 & -0.1 &  1.2\\
		P-2000-7  &  0.5 &  1.1 &  1.4 &  1.2 &  3.1\\
		P-2000-14 &  0.8 &  2.0 &  3.0 &  3.2 &  6.5\\
		P-2000-21 &  1.0 &  2.5 &  4.0 &  4.6 &  9.7\\
		\hline
		P-1000-1  & -0.1 &  0.0 & -0.1 & -0.2 &  2.3\\
		P-1000-7  &  0.6 &  1.4 &  1.6 &  1.5 &  5.0\\
		P-1000-14 &  0.9 &  2.4 &  3.1 &  3.5 &  9.0\\
		P-1000-21 &  1.3 &  3.4 &  4.6 &  5.3 & 13.1\\
		\hline
		\hline
		\multicolumn{6}{r}{ }\\
		\multicolumn{6}{r}{Fluctuating noise (ICRA5-250)}\\
		\hline
		\hline
		Profile & \footnotesize \footnotesize PLATT-2 & \footnotesize PLATT-4 & \footnotesize PLATT-6 & \footnotesize PLATT-8 & NH\\
		\hline
		P-8000-1  & -0.0 & -0.5 & -1.8 & -2.0 &  0.0\\
		P-8000-7  &  1.2 &  1.9 &  1.9 &  1.4 &  4.6\\
		P-8000-14 &  1.2 &  2.9 &  4.2 &  4.3 &  8.7\\
		P-8000-21 &  1.8 &  4.7 &  5.7 &  5.9 & 12.5\\
		\hline
		P-4000-1  &  0.4 & -0.7 & -0.8 & -1.0 &  3.2\\
		P-4000-7  &  1.3 &  2.7 &  2.0 &  2.4 &  7.2\\
		P-4000-14 &  2.1 &  3.7 &  4.6 &  4.8 & 11.4\\
		P-4000-21 &  2.0 &  4.7 &  6.1 &  7.0 & 14.7\\
		\hline
		P-2000-1  &  0.6 &  0.4 &  0.0 &  0.2 &  6.5\\
		P-2000-7  &  1.3 &  2.5 &  2.2 &  2.9 &  9.7\\
		P-2000-14 &  1.8 &  4.2 &  5.3 &  5.7 & 14.3\\
		P-2000-21 &  1.9 &  4.6 &  6.6 &  7.4 & 17.9\\
		\hline
		P-1000-1  &  0.3 &  1.0 &  2.1 &  1.8 & 15.0\\
		P-1000-7  &  1.9 &  3.3 &  4.1 &  5.1 & 18.8\\
		P-1000-14 &  1.3 &  4.3 &  5.7 &  6.6 & 22.4\\
		P-1000-21 &  2.1 &  4.9 &  6.2 &  7.8 & 26.3\\
		\hline
		\hline
	\end{tabular}
\end{table}
The tables present the simulated benefits in SRT using PLATT with expansion factors 2, 4, 6, and 8, for all 16 considered listener profiles, sorted by the limit frequency.
In the last column, the benefit in SRT that corresponds to the normal-hearing (NH) profile is presented as an orientation to assess the proportion of the total class D loss that be compensated.
Positive values indicate an improvement, that is, a lower SRT.
An orientative value for the random error of the presented simulation data in Table~\ref{tab:1} is 1\,dB.

For listener profiles with a level uncertainty of 1\,dB (P-*-1), only small benefits were observed.
In the stationary noise condition, processing the signals with PLATT had no effect on the SRTs ($\leq0.4$\,dB).
In the fluctuating noise condition, small positive and negative benefits (min. -2\,dB and max. 2\,dB) were observed; the SRTs for the listener profile P-1000-1 improved (lower SRTs) and the SRTs for the listener profiles P-8000-1 and P-4000-1 increased (negative improvement).
This effect was most pronounced with PLATT-8 and PLATT-6, less pronounced with PLATT-4, and could not be observed with PLATT-2, that is, it depended on the expansion factor.
The processing with PLATT, by increasing the amplitude of certain spectral modulations, decreases the prominence of (or \enquote{masks}) other modulation patterns, and it introduces a possible leak of information across frequencies, e.g., from $>1000$\,Hz to $<1000\,Hz$.
While the masking of other modulation patterns could explain the detrimental effect of PLATT for the profiles P-8000-1 and P-4000-1, the leakage of information from higher frequencies could explain the improvement for profile P-1000-1.
Despite these small variations of the SRTs due to the processing with PLATT in the fluctuating noise condition, one might feel comfortable to state:
As expected, the effect of a limited frequency range (an increase in SRT that cannot be compensated by simple amplification) cannot be compensated with the expansion performed by PLATT.

The data in the column NH, that is, the difference in SRT between the simulation with a specific listener profile and the normal-hearing profile, shows again how different the simulation results are in the stationary and fluctuating noise condition.
While limiting the frequency range to 1000\,Hz (P-1000-1) only resulted in a surprisingly small increase in SRT, namely by 2.3\,dB, the same modification increased the SRT in the fluctuating noise condition by 15.0\,dB.
This reaffirms the use of different noise maskers to assess speech in noise recognition performance.

Interpreting the addition of maskers to a speech signal as a frequency-dependent removal of information, the results indicate that the information in the mid-frequency range (1000-4000\,Hz), was much more relevant in the considered fluctuating noise condition (ICRA5-250) than in the stationary noise condition (OLNOISE).
In other words, in the stationary noise condition, this mid-frequency portion was mostly redundant at the SRT, and the ASR system could discriminate the 50 words of the matrix sentence test almost equally well only using the low-frequency information.
This was not the case for the fluctuating noise condition, where the mid-frequency portions contributed a substantial part of the information to achieve low SRTs (say, less than -15\,dB).
For the following presentation of the benefits with PLATT, it is important to keep in mind that this missing information, by design, cannot be compensated with PLATT.
That means, when considering the maximum achievable improvement for a specific profile, the increase in SRT due to limiting the frequency range was disregarded.
For example, with profile P-2000-14 in the fluctuating noise condition, the difference to the normal-hearing SRT is 14.3\,dB; improving the SRT by 14.3\,dB the performance of the normal-hearing profile would be achieved.
When only considering the increase in SRT due to limiting the frequency range, that is profile P-2000-1, the difference to the normal-hearing SRT is 6.5\,dB.
Hence, the maximum achievable improvement for profile P-2000-14 is then estimated by the difference $14.3-6.5=7.8$\,dB.

For profiles with increased level uncertainty (P-*-7, P-*-14, and P-*-21), the average benefits in SRT due to using PLATT were positive and ranged from 0.5 to 7.8\,dB.
For these profiles, the lowest improvements were found with PLATT-2, and the highest improvements often, but not always, with PLATT-8.
Overall, in both noise conditions, a very strong relation between the increase in level uncertainty and the improvement in SRT due to using PLATT could be observed.
Combined, there was strong general tendency that with higher level uncertainty and with higher expansion factors larger improvements were observed.
Exceptions were observed only for the profiles with a level uncertainty of 7\,dB (P-*-7), where higher expansion factors did not further increase the improvement.
This data supports the sensible expectation that lower expansion factors are sufficient (or even optimal) to compensate lower values of level uncertainty.
For profiles with a level uncertainty of 14 or 21\,dB, the improvements always increased together with the expansion factor.
With high values of level uncertainty, higher expansion factors might further improve the SRT; however, increasing the dynamic of a signal portion eight fold might have undesirable collateral effects which are discussed later.

In absolute terms, the improvements were generally larger in the fluctuating noise condition than in the stationary noise condition.
For example with the extreme profile P-1000-21 and PLATT-8 compensation, the improvement was 5.3\,dB in the stationary noise condition, and 7.8\,dB in the fluctuating one.
However, relating the improvement to the performance with the normal-hearing profile (rightmost column in Table~\ref{tab:1}), 5.3\,dB of a total class D loss of 13.1\,dB were compensated in the stationary noise condition and \enquote{only} 7.8\,dB of a total class D loss of 26.3\,dB.
While this interpretation correctly reflects the proportion of the total class D loss that was compensated, it does not reflect that a part of the class D loss cannot be compensated by expansion approach with PLATT by design. 
As explained earlier, the portion of the class D loss due to limiting the frequency range to 1000\,Hz is very different in both maskers.
To evaluate the achieved improvements with respect to the maximum achievable improvement, the class D loss due to limiting the frequency range needs to be disregarded.
In the context of the maximum achievable improvement, PLATT-8 compensated 5.3\,dB of $(13.1-2.3=)10.8$\,dB and 7.8\,dB of $(26.3-15.0=)11.3$\,dB of the class D loss due to a level uncertainty of 21\,dB in the stationary and fluctuating noise condition, respectively.
For the intermediate mixed profile P-2000-14, PLATT-6 compensated 3.0\,dB of $(6.5-1.2=)5.3$\,dB and 5.3\,dB of $(14.3-6.5=)7.8$\,dB in the stationary and fluctuating noise condition, respectively.
And for the least extreme mixed profile P-4000-7, PLATT-4 compensated 1.0\,dB of $(1.5-{-0.1}=)1.6$\,dB, and 2.7\,dB of $(7.2-3.2=)4.0$\,dB.
Hence, in relative terms, over a broad range of assumed parameters, PLATT expansion compensated at least about half of the class D loss caused by an increase of a level uncertainty.

The absolute improvements due to PLATT tended to increase with an increased limitation of the frequency range.
The effect of the level uncertainty increased with an increasing limitation of the frequency range.
Based on the presented data, however, it is difficult to make statements about the frequency-dependency of an optimal expansion factor because of the diverse non-linear interactions between the considered parameters.

\subsection*{SNR-dependency of benefits in SRT}
\begin{figure}
	\centerline{\includegraphics[width=\columnwidth]{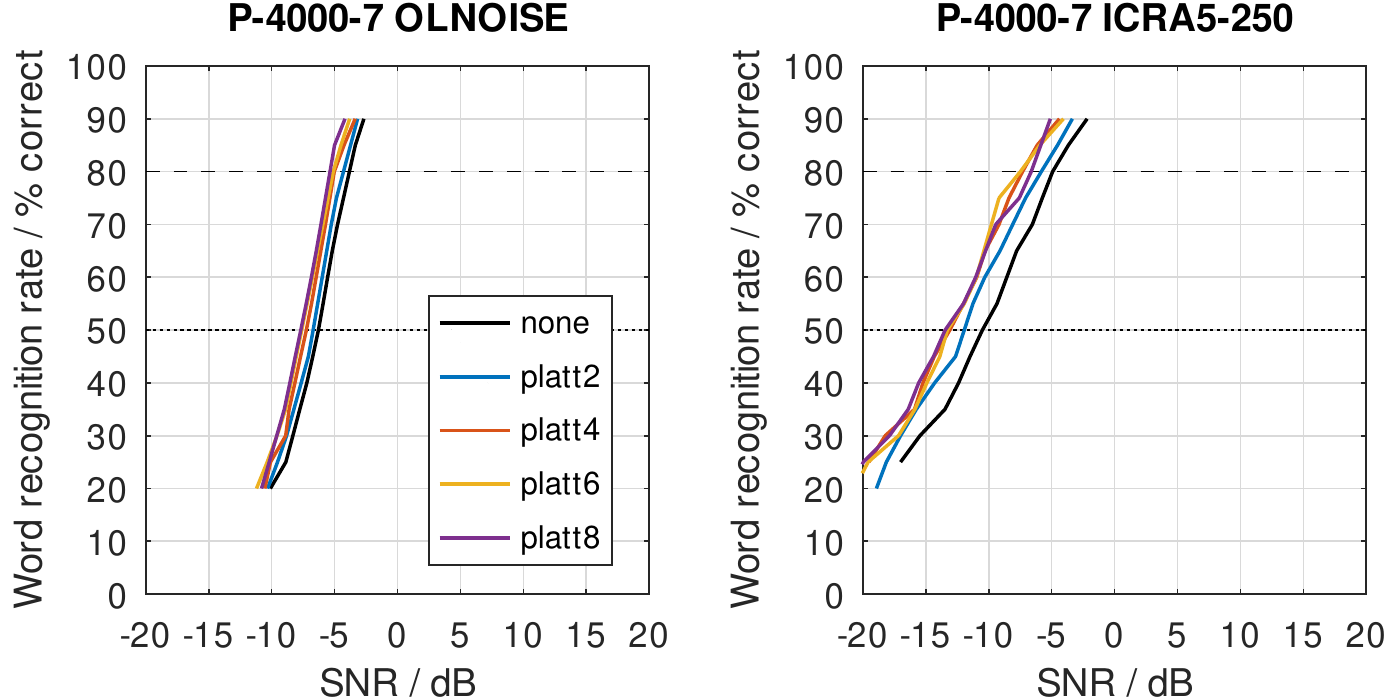}}
	\caption{Segments of psychometric functions for simulations with listener profile P-4000-7 without and with PLATT expansion factors 2, 4 and 6 in stationary noise (left panel) and fluctuating noise (right panel).
	The dotted and dashed lines indicate a word recognition rate of 50\% and 80\% correct, respectively.}
	\label{fig:19}
\end{figure}

\begin{figure}
	\centerline{\includegraphics[width=\columnwidth]{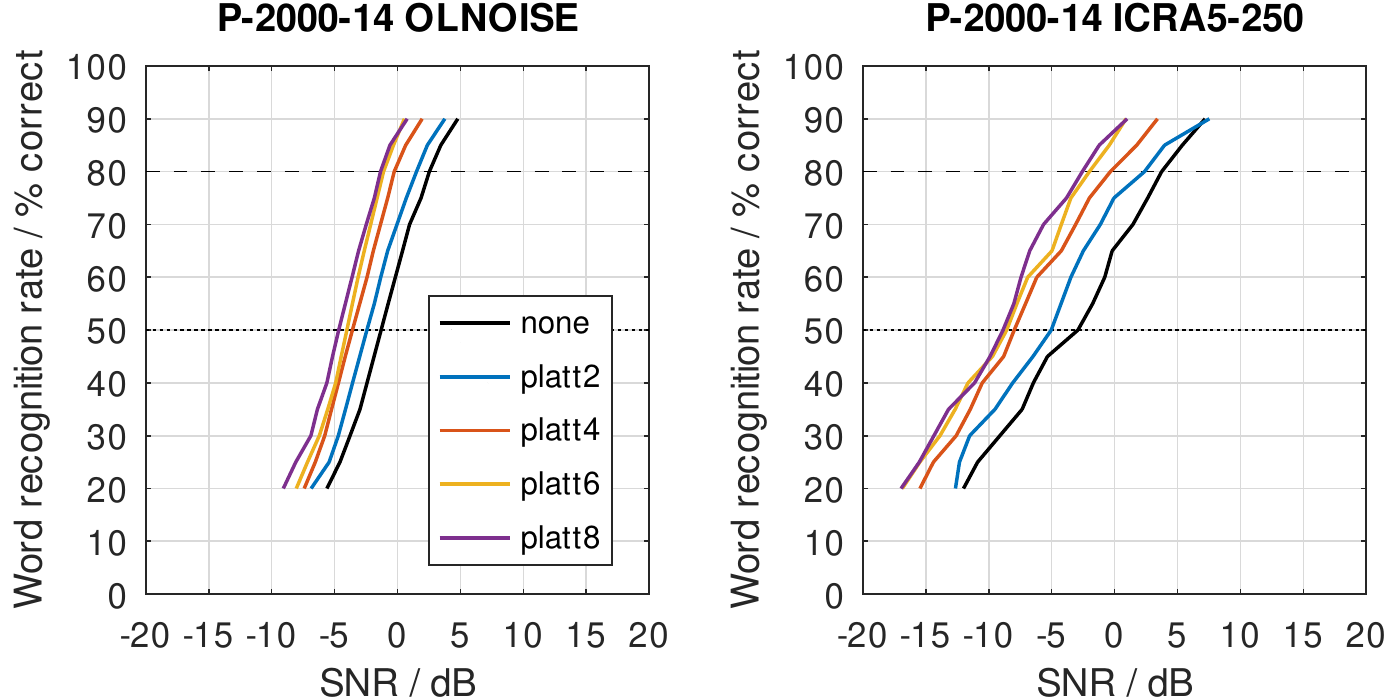}}
	\caption{Segments of psychometric functions for simulations with listener profile P-2000-14 without and with PLATT. Analog to Figure~\ref{fig:19}.}
	\label{fig:20}
\end{figure}

\begin{figure}
	\centerline{\includegraphics[width=\columnwidth]{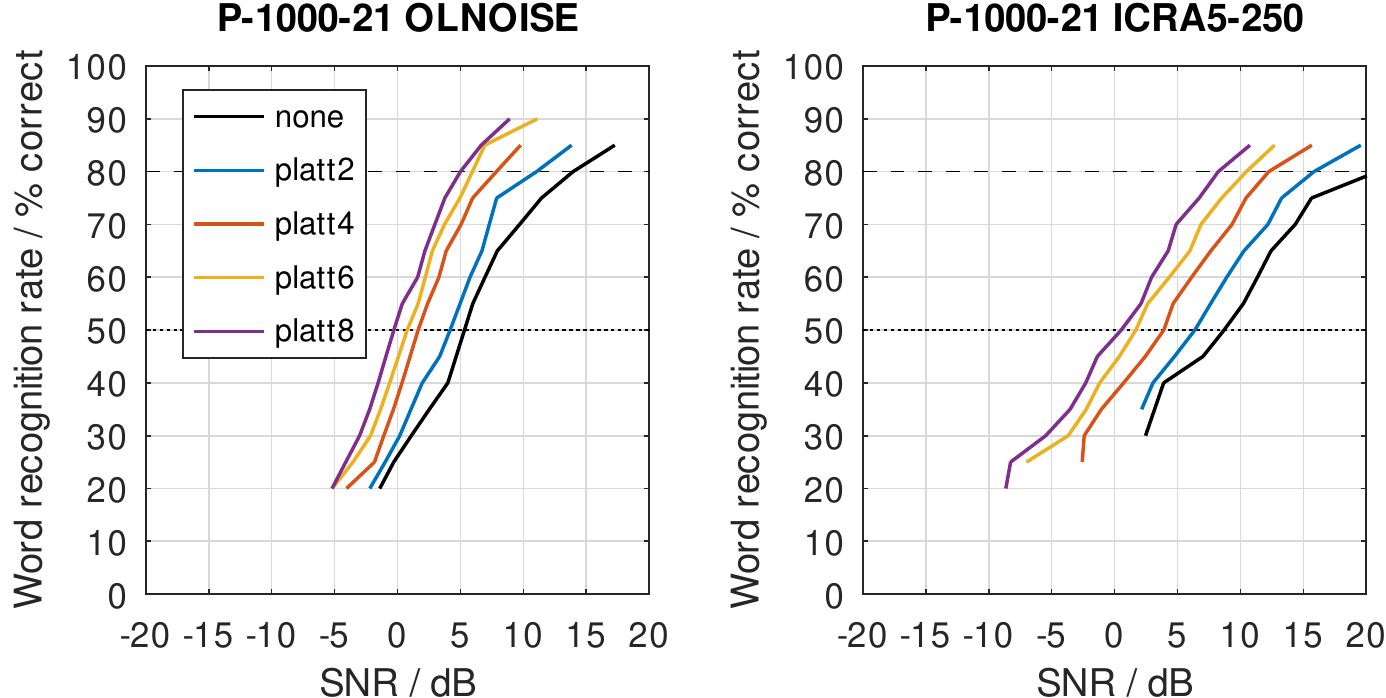}}
	\caption{Segments of psychometric functions for simulations with listener profile P-1000-21 without and with PLATT. Analog to Figure~\ref{fig:19}.}
	\label{fig:21}
\end{figure}

We (humans) generally prefer to have conversations at higher SNRs than the SRT-50, that is, at SNRs at which more than 50\% of the words can be correctly recognized; which might correspond to something like the SRT-80.
The SRT-80 can be simulated, but it cannot be as efficiently and accurately measured in listening experiments as the SRT-50, because of the shallower slope of the psychometric function at the SRT-80.
The main reason for simulating the SRT-50 was that it can be accurately measured in later listening experiments with human listeners, and accurate measurements are a requirement to show effects as small as 1\,dB.
To assess if the improvements would (at least according to the model) translate to improvements at SRTs preferred in real conversations, the psychometric functions of aided and unaided conditions were compared.
Segments of psychometric functions were obtained by evaluating simulations at SRT-20, SRT-25, ..., SRT-90.
Figure~\ref{fig:19}, \ref{fig:20}, and \ref{fig:21} present the unaided and aided psychometric functions for the mixed profiles P-4000-7, P-2000-14, and P-1000-21, respectively.
The simulation results in the unaided conditions are plotted in black, the corresponding simulation results with PLATT2, PLATT4, PLATT6, and PLATT8 expansion in blue, red, yellow, and purple color, respectively.
As expected, the slopes in the fluctuating noise conditions (right panels) were shallower than the slopes in the stationary noise condition (left panel).
Also, as expected, the data points in the fluctuating noise conditions were more noisy, due to the greater variability in the spectro-temporal distribution of the masker energy.
This variability could be decreased by increasing the amount of training data and testing data.
Within the uncertainty due to this variability, no reduction in improvement between SRT-50 and SRT-80 can be observed for listener profile P-4000-7.
For profile P-2000-14 (in Figure~\ref{fig:20}), where the improvements are larger, there was a trend towards a slight increase in slope with higher expansion factors.
This indicates, that the improvements due to the expansion with PLATT of the SRT-80 might be slightly larger than the corresponding improvement of the SRT-50.
This trend was confirmed by the data with the profile P-1000-21 in Figure~\ref{fig:21}.
According to the these simulations, the expansion with PLATT was found to improve the simulated SRT-80 to the same extent or even more than the SRT-50.

\section*{Discussion}
\label{sec:discussion}
The presented experimental results were derived using a model of auditory perception, more specifically, a model of impaired human speech recognition based on automatic speech recognition.
The modeling approach with FADE brings assumptions about the impaired human speech recognition process into a form in which they can be tested by comparing predictions with empirical data.
The employed model, the framework for auditory discrimination experiments (FADE) such as it was used by \cite{schaedler2020a}, was already evaluated with respect to predictions of the individual aided speech recognition performance of listeners with impaired hearing.
There, as elaborated in the \nameref{sec:intoduction}, an important assumption was that the part of the hearing loss which cannot be explained by the absolute hearing threshold, that is, the missing piece to describe the effect of hearing loss on speech recognition in noise, can be explained by the level uncertainty.
Another assumption was, that this model parameter also affects tone in noise perception and hence its value could be inferred from tone in noise detection tests.
While the results supported this hypothesis, evidence was not sufficient to rule out other mechanisms that would also increase the SRTs in noise.
This is a fundamental problem in modeling the individual speech recognition performance.
While the quantity that is predicted by the model, the SRT, can be measured in experiments with human listeners, the outcome of such measurements still depends on many correlated and non-linearly interacting parameters; some of which cannot be controlled very well, such as, e.g., attention.
And even if the experimental results were measured with the most accurate methods, the measurement errors include this uncontrollable (human) variability which can increase the required amount of data to falsify hypotheses to infeasible regions.
This is especially true for hypotheses which predict relatively small effects.
Hence, there is reasonable doubt about whether the removal of information in listeners with impaired hearing is really well described by the level uncertainty, or if it just coincidentally increased the SRT in the correct conditions.

To more specifically test if the level uncertainty is suitable to describe the effect of hearing loss on speech recognition in noise, a promising approach is to interact with it.
The expansion of PLATT was specifically designed to interact with---namely compensate---the effect of the level uncertainty.
The modeled data clearly showed this interaction.
If this specific interaction was found in empirical data, it would be strongly supportive for the hypothesis of the existence of a mechanism in the human auditory system similar to the level uncertainty.
Beyond the academical interest in the suitability of the assumptions to describe impaired human speech recognition performance, a positive result would have immediate practical implications for the design of hearing loss compensation strategies.

Let's remember that the goal was not to test if the expansion with PLATT improves speech recognition performance of listeners with impaired hearing.
For that, measurements with listeners with impaired hearing will be necessary.
The aim of this contribution was to:
\begin{itemize}
	\item[A)] Present an approach which is able to partially compensate a class D loss as implemented with the level uncertainty in FADE, and
	\item[B)] objectively evaluate this approach and come up with testable quantitative hypothesis on the benefit in noisy listening conditions.
\end{itemize}
The latter aim, in other words, was to guide the planning of the measurements with listeners with impaired hearing towards optimal evidence.
Hence, the following discussion is oriented towards the planning of a suitable experiment.

\subsection*{What is a realistic class D loss?}
\cite{plomp1978} did not distinguish mechanisms causing the postulated class D hearing loss which he used to describe hearing loss in noise.
His D component described the total loss in SRT in noise due to impaired hearing, also referred to as speech hearing loss D (SHL$_\text{D}$).
He reported average empirical values of SHL$_\text{D}$ from five investigations which range from as low as 1\,dB to above 10\,dB (cf. TABLE~III in \cite{plomp1978}).
In stationary noises (average speech spectrum noise, white noise, and airplane cockpit noise) the maximum average values for SHL$_\text{D}$ were 6 to 8\,dB.
In fluctuating noises (interfering talker and voice babble) the maximum average values were larger, up to 14\,dB.
The increase in SRT in noise was related to the increase in SRT in quiet SHL$_\text{A+D}$ in FIG.~8 in \cite{plomp1978}, which showed that every 3-dB increase of SHL$_\text{A+D}$ comes with a 1-dB increase of SHL$_\text{D}$.
This data indicated that on average, about a third of the hearing loss in quiet (SHL$_\text{A+D}$) was incompensable by simple amplification; that would be \emph{a lot}.
Listeners with speech hearing loss in quiet of 39\,dB would have on average 13\,dB of traditionally incompensible loss.
Also, \cite{plomp1978} observed that individual data differed considerably from the mean.
This individual variability could be due to individually different degrees of and causes for the measured loss, or due to insufficient measurement accuracy.
An important observation he made was that data points from listeners with age-related hearing loss agreed well with data points based from studies on other sensorineural hearing impairments.
As a consequence he considered age-related hearing loss to be primarily due to deterioration in the auditory pathway rather than to mental impairment.
This last point is fundamental considering that mental impairment can most likely not be compensated by signal processing.

However, these findings have to be taken with care.
The data used by \cite{plomp1978} were measured at different sites with different speech tests using different setups.
The resulting list of possible systematic and random errors in the underlying data is long.
The main contribution to the systematic error (expectable differences across studies) apart from the calibration error and the different listener panels was probably the use of different speech tests (including measurement paradigm, speech material, masker, presentation, ...).
It is known, that the type of speech material (e.g., logatomes, numbers, isolated words, or sentences) makes a difference in the outcome of a speech in noise recognition experiment.
Hence, tests with different speech material might also be differently susceptible to hearing loss and result in different values of SHL$_\text{D}$.
There is no reason to assume that SHL$_\text{D}$ is independent from the speech test.
The main contribution to the random error (unpredictable differences across measurements) was probably different for the studies and due to the stochastic nature of the measurement procedures.
Both, the random and the systematic errors were not specifically considered in the analysis of \cite{plomp1978}.
The variety of unknowns makes it difficult to translate his finding to expectable values and individual variability of SHL$_\text{D}$ with the employed matrix sentence test.
Matrix sentence tests were designed to minimize the random error in the measurement of an SRT, where \cite{kollmeier2015} reported a test-retest reliability of 0.5\,dB for listeners with normal hearing and 0.7\,dB for listeners with impaired hearing.

Fortunately, a suitable data set was measured with a matrix sentence test.
\cite{wardenga2015} found a similar relation between the degree of hearing loss and SHL$_\text{D}$ for the German matrix sentence test in the (stationary) test-specific noise condition, where the SRT in noise increased by about 1\,dB every 10-dB increase in PTA\footnote{pure-tone average} for PTAs below 47\,dB~HL (cf. Figure~5 there).
This is much less than the average (1\,dB every 3-dB) increase found by \cite{plomp1978}.
One reason for the lower increase could be that the speech hearing loss in quiet (SHL$_\text{A+D}$) includes the speech hearing loss in noise (SHL$_\text{D}$), while the PTA might not.
But that alone cannot explain the huge difference.
Another reason for the difference could be that \cite{wardenga2015} derived the relation from individual data with relatively low PTAs while \cite{plomp1978} derived the relation from averaged including all degrees of hearing impairment.
It is probably fair to assume that the data from \cite{wardenga2015} offers a more conservative and accurate estimate of the minimum expected average speech hearing loss in noise.

Taking the relation from \cite{wardenga2015} (which is based on data from 177 listeners aged 17 to 82y) as an orientation, it indicates that the average SHL$_\text{D}$ for PTAs of 60\,dB~HL could be about 6\,dB in stationary noise.
The individual variability of SHL$_\text{D}$ in their analysis was reported with 1.17\,dB for PTAs below 47\,dB~HL, which was only twice the test-retest reliability of the matrix test.
An interpretation of this relation could be that a given hearing loss in quiet is very likely related to a certain hearing loss in noise.
This interpretation would be only valid if additional amplification would not change the relation, that is, if the same relation was observed for a noise presentation level of, e.g., 75\,dB SPL instead of 65\,dB SPL.
However, it is difficult to speculate on that.
On the one hand, the test-specific noise (OLNOISE) reduces the effect of the individual hearing threshold by masking the speech signals with a stationary matched-spectrum noise.
On the other hand, for higher PTAs, the individual hearing threshold will eventually exceed the noise level at high frequencies, which could be compensated by amplification.
Then again, according to the simulations presented in this contribution, the removal of high-frequency portions ($>4000$\,Hz) of the signals in the OLNOISE condition had little effect on the SRT.
These considerations demonstrate the inherently non-linear relations between the parameters assumed to affect speech recognition in noise, and hence the difficulty of abstracting these from a given context.

Nonetheless, to continue with an specific value for SHL$_\text{D}$ for the German matrix sentence test that could likely be found in the wild, the relation established by \cite{wardenga2015} is assumed to ve valid.
In favor of this assumption is that the linear regression estimated from the data with PTAs below 47\,dB~HL describes the data in their Figure~5 (solid black line) really well.
One would expect an increase in variability with the PTA (below 47\,dB~HL) if the absolute hearing threshold affected the SRT, such as it was observed for higher PTAs; but this was not the case.
Accordingly, \cite{wardenga2015} concluded that with 65~dB SPL fixed noise presentation level the SRT is determined by listening in noise for PTAs $<\approx47$\,dB~HL, and above it is determined by listening in quiet.
Based on these considerations, it seems likely that listeners with an SRT of 0\,dB in the German matrix sentence test in the test-specific stationary noise exist whose speech recognition performance cannot be improved further by simple amplification.
That would indicate an SHL$_\text{D}$ of about 7\,dB.

This estimate can be used as an orientation to interpret the right-most column in Table~\ref{tab:1}, which is equivalent to the SHL$_\text{D}$.
For example, profile P-2000-14 with an SHL$_\text{D}$ of 6.5\,dB lies within the range of empirically observed values.
In this context, profiles P-1000-14 and P-2000-21 might be border cases with values of 9 to 10\,dB for SHL$_\text{D}$, and profile P-1000-21 lies probably outside range of empirically observed values for SHL$_\text{D}$.
Profiles P-8000-21, P-4000-14, P-2000-14, and P-1000-7, are all compatible with a large SHL$_\text{D}$ of less than 7\,dB in the stationary noise condition.
The key difference between these profiles is the proportion of SHL$_\text{D}$ that is due to the level uncertainty and hence might be compensable by a PLATT expansion. 
Assuming that the average available frequency range to listeners with impaired hearing is between 2000 and 4000\,Hz, P-4000-14 and P-2000-14 could represent parameter values in a realistic range.

The aim of classifying profiles into more (and less) realistic ones is to use them to formulate a quantitative hypothesis on the expected benefit in SRT due to PLATT expansion.
This does not mean that the considered profiles exist.
The assumption that the reduction in speech recognition performance in noise is due to a combination of a limited frequency range and a mechanism similar to an increased level uncertainty justifies their use.
While this assumption may be sensible, it doesn't mean that it is correct.
This would have to be tested in listening experiments.
For the effect of the limited frequency range, this could be achieved by, e.g., low-pass filtering the signals in speech recognition experiments.
For the effect of an increased level uncertainty, a direct verification is currently not possible because the level uncertainty adds a noise in a domain which is not accessible for manipulations in experiments with human listeners; unlike for the limitation of the frequency range, there is no known equivalent signal manipulation.
As explained in the \nameref{sec:intoduction}, the most promising (and at the same time constructive) approach to test whether a mechanism similar to an increased level uncertainty causes a part of the class D hearing loss of listeners with impaired hearing is trying to compensate it.

\subsection*{Role of PLATT expansion}
The expansion feature of the PLATT dynamic range manipulation approach aims to mitigate the increase in SRT due to an increased level uncertainty.
It was specifically designed to compensate the effect of the level uncertainty on speech recognition performance such as it is implemented in FADE.
The presented simulation results showed that this necessary interim goal was achieved and indicate that about half of class D loss due to an increased level uncertainty was compensated.
This only demonstrates that the expansion works as intended in the context of the model.
There is no reason to assume that the results, that is, the partial compensation of a class D hearing loss, would transfer to experiments human listeners.
Now---the key question of this contribution---what will happen in experiments with human listeners, in case the central model assumptions are incorrect?

\begin{itemize}
	\item[A)] If the PLATT expansion does not improve the SRTs in noise of human listeners with a class D hearing loss, this would indicate that the mechanism that causes the class D loss is not well described by the level uncertainty, because the PLATT expansion interacts differently with the mechanism in human listeners than with the mechanism in the model.
	\item[B)] If the PLATT expansion improves the SRTs in noise of human listeners and if these improvements are in line with individual predictions, this would strongly support the hypothesis that a mechanism similar to an increased level uncertainty causes a part of the class D hearing loss of listeners with impaired hearing.
	\item[C)] However, more likely than B), if the PLATT expansion improves the SRTs in noise of human listeners but the improvements are found to be substantially smaller than predicted, this would only partly support the hypothesis that a mechanism similar to an increased level uncertainty causes a part of the class D hearing loss of listeners with impaired hearing.
\end{itemize}
In scenario A), a central model assumption is incorrect.
The (to-be) collected empirical data still can help to improve the model, but the expansion feature of PLATT would be useless in the context of a hearing device.
In scenario B), the central model assumptions are probably correct.
The (to-be) collected empirical data cannot help to improve the model, but the expansion feature of PLATT would overcome a serious limitation of current hearing device technology.
In scenario C), most of the central model assumptions were probably correct.
The (to-be) collected empirical data can probably help to improve the model, and the expansion feature of PLATT could give strong hints regarding how to overcome a serious limitation of current hearing device technology.
In this last scenario, a differentiated analysis of the predictions errors will be helpful in tracking down the inaccurate assumptions.
The probabilities for scenario A), B), and C) are unknown.
To provide evidence for any scenario, an experiment with human listeners suitable to unmistakably show the compensation effect has to be conceived.

\subsection*{Potential of compensating a class D loss}
The presented simulation results clearly show the potential of the expansion of spectral modulation patterns in the range between 2 and 4 ERB, as implemented in PLATT, to compensate a class D hearing loss that was implemented by an increased level uncertainty.
In both noise conditions, approximately half of the class D loss due to an increased level uncertainty was compensated, while the class D loss due to a limited frequency range was not compensated.
Narrowing down the simulation results to profiles P-4000-14 and P-2000-14, which were identified earlier as more realistic, the predicted improvements in SRT with PLATT-6 were 3.0\,dB in the stationary noise condition.
These improvements would correspond to a compensation of $\frac{3.0}{4.9}\approx61\%$ and $\frac{3.0}{6.5-1.2}\approx57\%$ of the class D hearing loss due to the increased level uncertainty.
In the fluctuating noise condition, improvements in SRT of 4.6 and 5.3\,dB were predicted with PLATT-6, with profiles P-4000-14 and P-2000-14, respectively.
These improvements would correspond to a compensation of $\frac{4.6}{11.4-3.2}\approx56\%$ and $\frac{5.3}{14.3-6.5}\approx68\%$ of the class D hearing loss due to the increased level uncertainty.
Hence, if the model assumptions are correct, benefits in SRT of about 3.0\,dB and 5.0\,dB in the stationary and fluctuating noise condition are expected, which correspond to more than 50\% of the respective class D losses due to the level uncertainty.

Such benefits would be individually measurable with the standard German matrix sentence test.
However, according to the profile P-4000-7, for which benefits in SRT of up to 1.3\,dB and 2.7\,dB were predicted in the stationary and fluctuating noise condition, respectively, individual measurements would possibly not yield significant results.
This is because the individual benefit in SRT is derived from two individual measurements (aided and unaided), each generating a random error of about 0.7\,dB (standard deviation) for listeners with impaired hearing.
If the real benefit is lower than about ($0.7\cdot\sqrt{2}\cdot1.96\approx1.94\approx$)2\,dB, a significant (with p-value $<0.05$) effect can be only shown in group averages but not in single individual measurements.
However, showing an effect in individual measurements is highly preferable because then the data could be used to further analyze the individual characteristics of listeners with and without benefits.
Hence, even if this goal might not be achieved, experiments with human listeners should be designed with the aim to maximize the absolute effect.

\subsection*{Optimal values for the expansion}
The optimal (in terms of improvements in SRT) value for the expansion factor depended on the level uncertainty parameter.
For lower values of the level uncertainty, in many conditions lower factors (4 or 6) resulted in the best speech recognition performance, but higher values did not decrease the improvement much.
For the profiles P-4000-14 and P-2000-14, a factor of 8 was optimal.
From this perspective, nothing speaks against evaluating PLATT with high expansion factors in listening experiments.
Because an expansion factor of 8 results in a strong modification of the signal, and because it is unknown how such modifications are perceived in terms of quality and loudness by listeners with impaired hearing, at least two values for the expansion factor should be evaluated in listening experiments.

\subsection*{Audio quality and loudness}
\begin{figure*}[h!]
	\centerline{\includegraphics[width=1.0\textwidth]{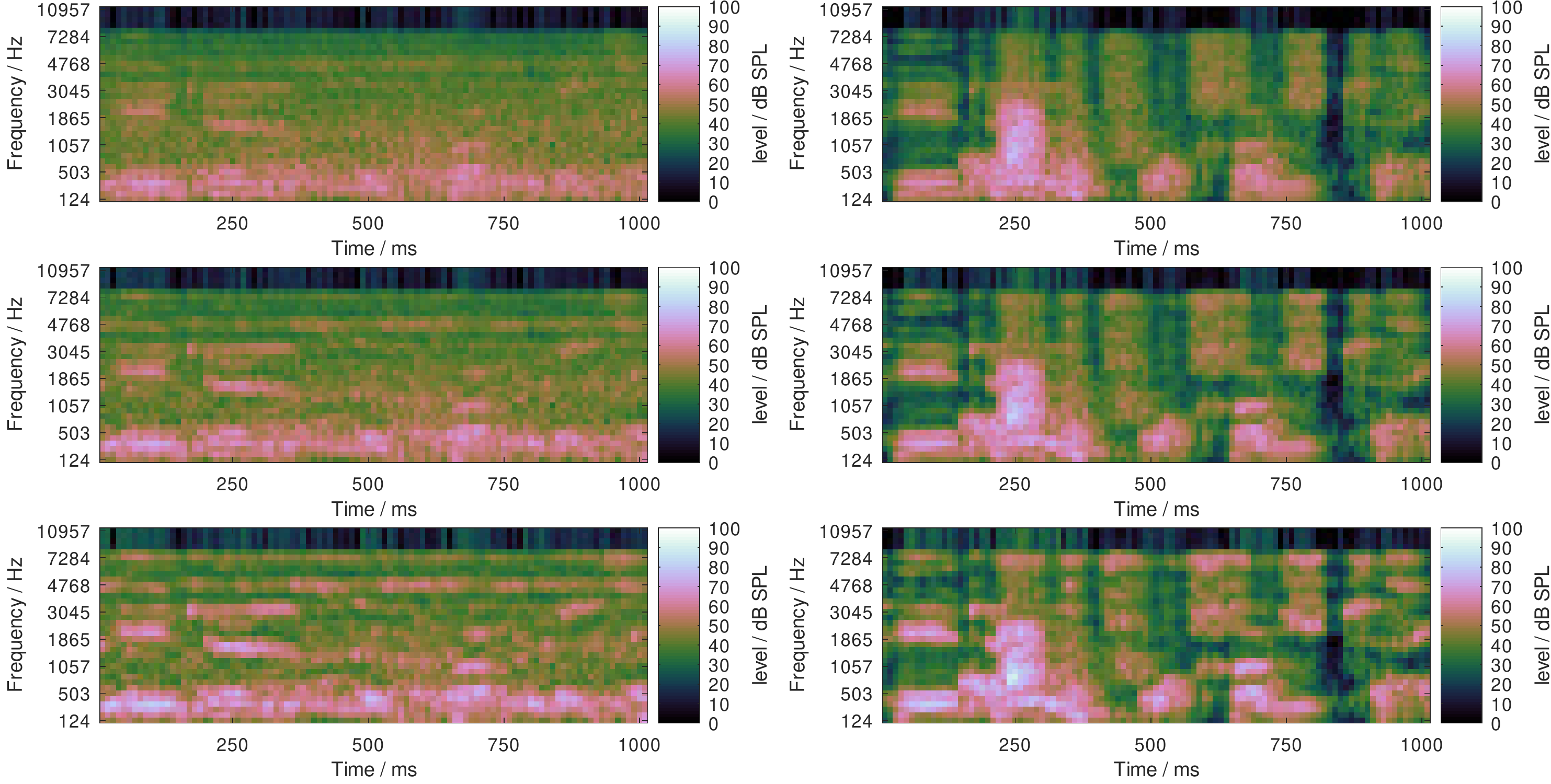}}
	\caption{Illustration of the effect of the expansion factor with PLATT: Log Mel-spectrograms of processed noisy speech in stationary (left column) and fluctuating noise (right column) at 0\,dB SNR processed with PLATT-1 (top row), PLATT-4 (center row), and PLATT-8 (bottom row).}
	\label{fig:22}
\end{figure*}

In measurements with human listeners, speech recognition performance, audio quality perception, and loudness perception all depend on the presentation level.
In this contribution, efforts were made to separate the speech recognition performance from the presentation level as much as possible.
This resulted in considering elevated presentation levels, and loudness perception is strongly related to presentation levels.
Also in this contribution, efforts were made to mitigate audible artifacts in the manipulation and resynthesis stages of PLATT.
But the non-linear manipulation inevitable results in audible artifacts which probably affect audio quality perception.
The effect of processing noisy speech signals at 0\,dB SNR with PLATT-1, PLATT-4, and PLATT-8 is illustrated in Figure~\ref{fig:22}.
The spectro-temporal maxima of the shown log Mel-spectrograms are 68.8, 73.4, and 79.8\,dB SPL for the noisy signals in stationary noise, and 77.0, 81.8, and 89.3\,dB for the noisy signals in fluctuating noise.
The corresponding effective amplitudes (RMS) are 66.6, 69.2, and 74.4\,dB SPL for the noisy signals in stationary noise, and 69.7, 72.3, 77.8\,dB for the noisy signals in fluctuating noise.
The increase in RMS and maximum power, with PLATT-8 by about 8\,dB and more than 10\,dB, respectively, suggests an increased loudness perception and also demonstrates the necessity to evaluate the approach in conditions in which an increase in level must not result in an improvement in SRT.
Also, the processing will probably have an effect on the perceived audio quality, where, with increased expansion factors, the audio quality will eventually decrease.
In the subjective opinion of the author, based on a comparison of unprocessed and processed signals, the artifacts due to processing the stimuli with PLATT-1 are minor, with PLATT-4 clearly audible, and with PLATT-8 pronounced.
While it would be a difficult task to predict the effect of the processing on the perception of audio quality of listeners with normal hearing, it would be even more so for listener with impaired hearing.

Apart from speech recognition performance, individual loudness and audio quality perception are important perceptual dimensions for hearing aid users.
The expected increase in perceived loudness and a possible decrease in perceived audio quality are considered collateral side-effects of a signal manipulation with PLATT.
These side-effects can either be tolerated or mitigated/compensated with further extensions or modifications of the PLATT approach.
It is unclear how a listener with impaired hearing would trade speech recognition performance for audio quality or loudness in the context of PLATT expansion.
To gather evidence about the relation of PLATT expansion, loudness perception, and audio quality perception, the considered expansion factors should cover a wide range of these side-effects.
Hence an evaluation of speech recognition performance, audio quality, and loudness with PLATT expansion factors of 1, 4, and 8 is proposed.
This will help to better estimate the suitability of the raw approach in the context of hearing aids and show which properties should be in the focus for possible future optimization.

There are several possibilities to optimize the expansion approach with respect to loudness perception and quality.
For example, the expansion factor can be chosen to be frequency dependent.
If the frequency regions in which the benefits are achieved by expansion do not fully overlap with the frequency regions where loudness perception is critical, there could be potential to favorably trade speech recognition performance against loudness.
The expansion factor could be limited, or mapped otherwise non-linearly to only expand (the low) modulation amplitudes required to improve speech perception in low-SNR conditions while leaving (the already high) modulation amplitudes in high-SNR conditions unmodified.
A detailed discussion of possible modifications is considered too hypothetical to be further elaborated in this contribution.

\subsection*{Proposed measurement conditions}
Based on the presented simulations and considerations, the following experimental measurement conditions are proposed to test the presented hypothesis that a part of a class D hearing loss can be compensated by expanding spectral patterns in the range between 2 and 4\,ERB in speech in noise recognition experiments with human listeners.
Regarding the noise maskers, an evaluation with the test-specific noise (e.g. OLNOISE for the German matrix sentence test), the fluctuating ICRA5-250 noise, and, in addition, a competing voice masker, e.g., the International Speech Test Signal \citep[ISTS;][]{holube2010} or its optimized version with limited pause durations, the International Female Fluctuating Masker (IFFM\footnote{\url{https://www.ehima.com/documents/}}), is proposed.
The IFFM masker is interesting because it shares the speech modulation properties with the target speaker and usually results in very weak masking for listeners with normal hearing.
It was not included in the objective evaluation with FADE because predictions with FADE for competing talker scenarios are known to be inaccurate; \cite{schaedler2018} reported the effect of the masker to be overestimated by approximately 10\,dB.

The noise presentation levels ideally would be sufficiently high to compensate any hearing loss which is compensable by simple amplification.
For the stationary noise condition, this would be easier to achieve than with the fluctuating noise conditions.
As shown in left panel of Figure~\ref{fig:2}, even mild hearing loss profiles are expected to require very high presentation levels to maximize the compensation with simple linear amplification.
However, the improvements in SRT at high presentation levels can be expected to be reasonably small.
As a compromise, and to avoid measuring the whole Plomp-curve, a reference experiment with an increased level can be performed to test if the equivalent linear amplification would result in a similar speech recognition performance than using PLATT expansion.
The observed increase in RMS level with PLATT-8 was about 8\,dB.
Adding a small margin, 10\,dB amplification should result in higher RMS levels of noisy speech signals at 0\,dB SNR than processing the same signals with PLATT-8.
Hence, noise presentation levels of 70 and 80\,dB SPL are proposed.
The proposed noise presentation level is 70\,dB SPL for aided and unaided measurements, while the noise presentation level of 80\,dB SPL serves as an alternative to the aided measurement; one in which 10\,dB amplification is applied.
If the processing with PLATT expansion achieves better SRTs than with the 10\,dB linear amplification, this would indicate the compensation of a class D hearing loss.
However, the RMS level is only a rough proxy to loudness perception.
Hence, in all conditions (70\,dB SPL unaided, 70\,dB SPL aided, 80\,dB SPL unaided), the loudness perception of characteristic signals, e.g., speech in noise at 0\,dB SNR, should be measured to check if the perceived loudness of the processed 70\,dB SPL signals is really lower than the perceived loudness of the 80\,dB SPL signals.

To further mitigate any benefits due to simple linear amplification at high presentation levels, all stimuli should be low-pass filtered to achieve an effective frequency range of 2000\,Hz.
This would limit the evaluation of a benefit due to PLATT expansion to the frequency range up to 2000\,Hz; the central frequency region that listeners with impaired hearing typically can use for speech recognition.
A focus on this frequency region could be advantageous in a first evaluation.
The low-pass filtering would make a group of listeners with impaired hearing less heterogeneous, because larger differences across listeners are usually observed at frequencies above 2000\,Hz. 
And the loudness perception of such low-pass filtered stimuli might also be more pleasant than their corresponding broadband variants.

In total, the proposed measurements comprise 18 conditions:
Three for training, three unaided measurements at 70\,dB SPL (one for each noise masker), three unaided measurements at 80\,dB SPL (also one for each noise masker), and nine measurements with PLATT-1, PLATT-4, and PLATT-8 (that is, three for each noise masker).
In all conditions, SRT, loudness perception and audio quality perception are recommended to be assessed.

In addition, it is recommendable to characterize the listeners in the way \cite{schaedler2020a} proposed it, that is, with tone detection and tone in noise detection experiments, and not only with clinical audiograms.
This additional data could be used to perform individual predictions of benefits with FADE and compare them to the individual measurements.
The comparison of individual measurements and predictions is crucial to 
detect invalid assumptions in the model.

\subsection*{Final practical considerations}
As speech material for the SRT measurements, the optimized matrix test sentences in native language are recommended \citep{kollmeier2015}.
These are phonetically balanced and optimized for a high test-retest reliability.
To minimize the training effect, three training lists of 20 sentences in noise are recommended.
Lists of 30 sentences obtain a higher test-retest reliability but can also result in excessively long measurements which can be fatiguing.
Hence, as a compromise, measurements with lists of 20 sentences could be used and repeated on a different day, which would also allow to detect a possibly remaining training effect.

For a first approach, headphone measurements are preferable over free-field measurements because they can be performed monaurally, which is recommendable.
A monaural presentation prevents the interference of possibly individual binaural effects and facilitates the comparison to model predictions, as discussed by \cite{schaedler2020a}.

\subsection*{Outlook}
Once there is evidence whether the proposed compensation strategy has a positive effect on speech recognition performance of listeners with impaired hearing, the band-width limitation to 2000\,Hz that was recommended for the first experiments should be removed.
For the next steps, it would not be important anymore to demonstrate the compensation of a class D loss alone, but to show benefits in more realistic and individually optimized configurations.
Hence, there should be a shift towards a joint individual optimization of the compensation of class A and class D loss, loudness perception, and audio quality.
Then, care should be taken to individually normalize loudness perception to get meaningful data.
If the expansion approach with PLATT proves to work in monaural listening conditions, the concept should be extended to a binaural listening condition.
For this, a free-field setup with a mobile hearing aid prototype hardware is recommendable to correctly assess individual binaural hearing, including binaural loudness perception.
To better understand which speech portions are most affected by PLATT expansion, it would also be interesting to study its effect on phonemic contrasts.

\section*{Conclusions}
\label{sec:conclusion}
The most important findings of this work can be summarized as follows:
\begin{itemize}
	\item The functional modeling of the class D hearing loss with the framework for auditory discrimination experiments (FADE), implemented by means of the level uncertainty, was interpreted as the counterpart of a compensation strategy which aims to (partially) compensate a class D heaing loss.
	\item The strict low-delay constraints in hearing aid applications only allow for a manipulation of mainly spectral modulation patterns. Of these, the patterns in the range of 2 to 4\,ERB seem especially suitable to protect them against the effect of the level uncertainty by dynamic range expansion.
	\item A low-delay, real-time capable implementation of a patented dynamic range manipulation scheme (PLATT) which allows to perform the required dynamic range expansion was proposed. The implementation was optimized to run in real-time on the Raspberry Pi 3 Model B platform.
	\item The evaluation of the PLATT expansion with FADE for several idealized profiles of hearing loss indicated that approximately half of the class D hearing loss due to an increased level uncertainty was compensable.
	\item Simulations with FADE were used to predict the outcomes of specific speech recognition experiments \emph{prior} to performing these. The underling hypothesis, that a class D hearing loss can be (partially) compensated, can be directly tested in an experiment with human listeners in the same listening conditions. Recommendations for this experiment were elaborated.
\end{itemize}

\bibliographystyle{apalike}

\begin{thebibliography}{}
	\bibitem[Bisgaard et~al., 2010]{bisgaard2010}
	Bisgaard, N., Vlaming, M.~S., and Dahlquist, M. (2010)
	\newblock Standard audiograms for the IEC 60118-15 measurement procedure.
	\newblock {\em Trends in amplification}, 14(2):113--120, \url{https://doi.org/10.1177%2F1084713810379609}

	\bibitem[Bustamante and Braida, 1987]{bustamante1987}
	Bustamante, D.~K. and Braida, L.~D. (1987)
	\newblock Principal-component amplitude compression for the hearing impaired.
	\newblock {\em The Journal of the Acoustical Society of America}, 82(4):1227--1242, \url{https://doi.org/10.1121/1.395259}

    \bibitem[Dreschler, 1992]{dreschler1992}
	Dreschler, W.~A. (1992)
	\newblock Fitting multichannel-compression hearing aids.
	\newblock {\em Audiology}, 31(3):121--131, \url{https://doi.org/10.3109/00206099209072907}

	\bibitem[Dreschler et~al., 2001]{dreschler2001}
	Dreschler, W.~A., Verschuure, H., Ludvigsen, C., and Westermann, S. (2001)
	\newblock ICRA noises: artificial noise signals with speech-like spectral and temporal properties for hearing instrument assessment.
	\newblock {\em Audiology}, 40(3):148--157, \url{https://doi.org/10.3109/00206090109073110}

	\bibitem[ETSI, 2007]{etsi2007}
	European Telecommunications Standards Institute (2007)
	\newblock "202 050 v1.1.5" Speech processing transmission and quality aspects (STQ); Distributed speech recognition; Advanced front-end feature extraction algorithm; Compression algorithms.
	\newblock {\em Standard}, \url{https://www.etsi.org/deliver/etsi_es/202000_202099/202050/01.01.05_60/es_202050v010105p.pdf}

	\bibitem[Grimm et~al., 2015]{grimm2015}
	Grimm, G., Herzke, T., Ewert, S., and Hohmann, V. (2015)
	\newblock Implementation and evaluation of an experimental hearing aid dynamic range compressor.
	\newblock In {\em Proceedings of German Annual Conference on Acoustics}, 185--188, \url{http://pub.dega-akustik.de/DAGA_2015/data/articles/000429.pdf}

	\bibitem[Hochmuth et~al., 2015]{hochmuth2015}
	Hochmuth, S., Kollmeier, B., Brand, T., and Jürgens, T. (2015)
	\newblock Influence of noise type on speech reception thresholds across four languages measured with matrix sentence tests.
	\newblock {\em International Journal of Audiology}, 54(sup2):62--70, \url{https://doi.org/10.3109/14992027.2015.1046502}
	
	\bibitem[Hohmann and Kollmeier, 1995]{hohmann1995}
	Hohmann, V. and Kollmeier, B. (1995)
	\newblock The effect of multichannel dynamic compression on speech intelligibility.
	\newblock {\em The Journal of the Acoustical Society of America}, 97(2):1191--1195, \url{https://doi.org/10.1121/1.413092}
	
	\bibitem[Holube et~al., 2010]{holube2010}
	Holube, I., Fredelake, S., Vlaming, M. and Kollmeier, B. (2010)
	\newblock Development and analysis of an international speech test signal (ISTS).
	\newblock {\em International Journal of Audiology}, 49(12):891--903, \url{https://doi.org/10.3109/14992027.2010.506889}
	
    \bibitem[Hülsmeier et~al., 2020]{huelsmeier2020}
    Hülsmeier, D., Warzybok, A., Kollmeier, B., and Schädler, M.~R. (2020)
	\newblock Simulations with FADE of the effect of impaired hearing on speech recognition performance cast doubt on the role of spectral resolution.
	\newblock {\em Hearing Research}, 395, \url{https://doi.org/10.1016/j.heares.2020.107995}
	
	\bibitem[ISO 226, 2003]{iso2003}
	ISO (2003).
	\newblock Standard 226: 2003: Acoustics--normal equal-loudness-level contours.
	\newblock {\em International Organization for Standardization}, 63, \url{https://www.iso.org/standard/34222.html}
	
	\bibitem[Kollmeier et~al., 2015]{kollmeier2015}
	Kollmeier, B., Warzybok, A., Hochmuth, S., Zokoll, M.~A., Uslar, V., Brand, T., and Wagener, K.~C. (2015)
	\newblock The multilingual matrix test: Principles, applications, and comparison across languages: A review.
	\newblock {\em International Journal of Audiology}, 54(sup2):3--16, \url{https://doi.org/10.3109/14992027.2015.1020971}.
	
	\bibitem[Kollmeier et~al., 2016]{kollmeier2016}
	Kollmeier, B., Schädler, M.~R., Warzybok, A., Meyer, B.~T., and Brand, T. (2016)
	\newblock Sentence recognition prediction for hearing-impaired listeners in	stationary and fluctuation noise with fade: Empowering the attenuation and	distortion concept by Plomp with a quantitative processing model.
	\newblock {\em Trends in Hearing}, 20, \url{https://doi.org/10.1177%2F2331216516655795}

	\bibitem[Levitt and Neuman, 1991]{levitt1991}
	Levitt, H. and Neuman, A.~C. (1991)
	\newblock Evaluation of orthogonal polynomial compression.
	\newblock {\em The Journal of the Acoustical Society of America}, 90(1):241--252, \url{https://doi.org/10.1121/1.401294}

	\bibitem[Moore et~al., 1999]{moore1999}
	Moore, B.~C.~J., Peters, R.~W., and Stone, M.~A. (1999)
	\newblock Benefits of linear amplification and multichannel compression for speech comprehension in backgrounds with spectral and temporal dips.
	\newblock {\em The Journal of the Acoustical Society of America}, 105(1):400--411, \url{https://doi.org/10.1121/1.424571}
	
	\bibitem[Plomp, 1978]{plomp1978}
	Plomp, R. (1978)
	\newblock Auditory handicap of hearing impairment and the limited benefit of hearing aids.
	\newblock {\em The Journal of the Acoustical Society of America}, 63(2):533--549, \url{https://doi.org/10.1121/1.381753}

	\bibitem[Plomp, 1988]{plomp1988}
	Plomp, R. (1988)
	\newblock The negative effect of amplitude compression in multichannel hearing aids in the light of the modulation-transfer function.
	\newblock {\em The Journal of the Acoustical Society of America}, 83(6):2322--2327, \url{https://doi.org/10.1121/1.396363}

	\bibitem[Schädler et~al., 2012]{schaedler2012}
	Schädler, M.~R., Meyer, B., and Kollmeier, B. (2012)
	\newblock Spectro-temporal modulation subspace-spanning filter bank features for robust automatic speech recognition.
	\newblock {\em The Journal of the Acoustical Society of America}, 131(5):4134--4151, \url{https://doi.org/10.1121/1.3699200}
	
	\bibitem[Schädler et~al., 2015]{schaedler2015}
	Schädler, M.~R., Warzybok, A., Hochmuth, S., and Kollmeier, B. (2015)
	\newblock Matrix sentence intelligibility prediction using an automatic speech recognition system.
	\newblock {\em International Journal of Audiology}, 54(sup2):100--107, \url{https://doi.org/10.3109/14992027.2015.1061708}
	
	\bibitem[Schädler et~al., 2016a]{schaedler2016a}
	Schädler, M.~R., Warzybok, A., Ewert, S.~D., and Kollmeier, B. (2016b)
	\newblock A simulation framework for auditory discrimination experiments: Revealing the importance of across-frequency processing in speech perception.
	\newblock {\em The Journal of the Acoustical Society of America}, 139(5):2708--2722, \url{https://doi.org/10.1121/1.4948772}
	
	\bibitem[Schädler et~al., 2016b]{schaedler2016b}
	Schädler, M.~R., Hülsmeier, D., Warzybok, A., Hochmuth, S., and Kollmeier, B. (2016a)
	\newblock Microscopic multilingual matrix test predictions using an ASR-based speech recognition model.
	\newblock In {\em Proceedings of INTERSPEECH}, 610--614, \url{http://dx.doi.org/10.21437/Interspeech.2016-1119}
	
	\bibitem[Schädler et~al., 2018]{schaedler2018}
	Schädler, M.~R., Warzybok, A., and Kollmeier, B. (2018)
	\newblock Objective Prediction of Hearing Aid Benefit Across Listener Groups Using Machine Learning: Speech Recognition Performance With Binaural Noise-Reduction Algorithms.
	\newblock {\em Trends in Hearing}, 22, \url{https://doi.org/10.1177/2331216518768954}.
		
	\bibitem[Schädler et~al., 2020a]{schaedler2020a}
	Schädler, M.~R., Hülsmeier, D., Warzybok, A., and Kollmeier, B. (2020)
	\newblock Individual Aided Speech-Recognition Performance and Predictions of Benefit for Listeners With Impaired Hearing Employing FADE.
	\newblock {\em Trends in Hearing}, 24, \url{https://doi.org/10.1177%2F2331216520938929}
	
	\bibitem[Schädler, 2020b]{schaedler2020b}
	Schädler, M.~R. (2020b)
	\newblock Optimization and evaluation of an intelligibility-improving signal processing approach (IISPA) for the Hurricane Challenge 2.0 with FADE.
	\newblock In {\em Proceedings of INTERSPEECH}, 1331--1335, \url{https://doi.org/10.21437/Interspeech.2020-0093}
  
	\bibitem[Souza, 2002]{souza2002}
	Souza, P.~E. (2002)
	\newblock Effects of compression on speech acoustics, intelligibility, and sound quality.
	\newblock {\em Trends in Amplification}, 6(4):131--165, \url{https://doi.org/10.1177%2F108471380200600402}

    \bibitem[Wagener et~al., 1999]{wagener1999}
    Wagener, K., Brand, T., and Kollmeier, B. (1999)
    \newblock Entwicklung und Evaluation eines Satztests für die Deutsche Sprache I-III: Design, Optimierung und Evaluation des Oldenburger Satztests.
    \newblock {\em Zeitschrift für Audiologie}, 38(1-3):4--15
  
    \bibitem[Wagener et~al., 2006]{wagener2006}
    Wagener, K.~C., Brand, T., and Kollmeier, B. (2006)
    \newblock The role of silent intervals for sentence intelligibility in fluctuating noise in hearing-impaired listeners.
    \newblock {\em International Journal of Audiology}, 45(1):26--33, \url{https://doi.org/10.1080/14992020500243851}
  
	\bibitem[Wardenga et~al., 2015]{wardenga2015}
	Wardenga, N., Batsoulis, C., Wagener, K.~C., Brand, T., Lenarz, T., and Maier, H. (2015)
	\newblock Do you hear the noise? The German matrix sentence test with a fixed noise level in subjects with normal hearing and hearing impairment.
	\newblock {\em International Journal of Audiology}, 54(sup2):71--79, \url{https://doi.org/10.3109/14992027.2015.1079929}.
  
  	\bibitem[Yund and Buckles, 1995]{yund1995}
	Yund, E.~W. and Buckles, K.~M. (1995)
	\newblock Multichannel compression hearing aids: Effect of number of channels on speech discrimination in noise.
	\newblock {\em The Journal of the Acoustical Society of America}, 97(2):1206--1223, \url{https://doi.org/10.1121/1.413093}
\end{thebibliography}

\end{document}